\documentclass[final]{IEEEtranTCOM}
\pdfoutput=1
\usepackage{amsmath,epsfig,amssymb,verbatim,amsopn,subfigure,color}
\usepackage{cite,xspace}
\usepackage{array,algorithm,algorithmic}
\usepackage{bbm}
\usepackage{array}
\usepackage{multirow}
\usepackage{multicol}
\usepackage{rotating}
\usepackage{dsfont,float}
\usepackage{stfloats}
\usepackage{enumerate,makecell}
\normalsize

\ifCLASSOPTIONpeerreview
\renewcommand{\baselinestretch}{1.69}
\fi

\ifCLASSOPTIONonecolumn
    
\else
    
\fi

\makeatletter
\renewcommand*{\@opargbegintheorem}[3]{\trivlist
  \item[\hskip \labelsep{\itshape #1\ #2}] {\itshape (#3):} {\normalfont}}
\makeatother

\newcommand{\E}{\Omega_{t_i}}
\newcommand{\A}{\frac{\beta}{\rho_0}}
\newcommand{\B}{\frac{\beta}{P_0r_0^{-\alpha}}}
\newcommand{\Bnew}{\beta r_0^{\alpha}}
\newcommand{\area}{\mathcal{A}}
\newcommand{\radius}{W}
\newcommand{\fr}{f_{R}(r)}
\newcommand{\fg}{f_{G}(g)}
\newcommand{\fgg}{f_{G_0}(g_0)}
\newcommand {\jj}{\mathbf{i}}
\usepackage{relsize}

\newtheorem{remark}{Remark}

\graphicspath{{figs/},{Figures/}}

\newcommand{\AuthorOne}{Jing~Guo, {\em{Student Member, IEEE}}}
\newcommand{\AuthorTwo}{Salman~Durrani, {\em{Senior Member, IEEE}}}
\newcommand{\AuthorThree}{Xiangyun~Zhou, {\em{Member, IEEE}}}
\newcommand{\ThankOne}{The authors are with the Research School of Engineering, College of Engineering and Computer Science, The
Australian National University, Canberra, ACT 0200, Australia.
Emails: \{jing.guo, salman.durrani, xiangyun.zhou\}@anu.edu.au.}

\begin{document}
\title{Outage Probability in Arbitrarily-Shaped Finite Wireless Networks}
\author{\authorblockN{\AuthorOne,~\AuthorTwo,~and \AuthorThree\thanks{\ThankOne}}}
\maketitle
%
\begin{abstract}
This paper analyzes the outage performance in finite wireless networks. Unlike most prior works, which either assumed a specific network shape or considered a special location of the reference receiver, we propose two general frameworks for analytically computing the outage probability at any arbitrary location of an arbitrarily-shaped finite wireless network: (i) a moment generating function-based framework which is based on the numerical inversion of the Laplace transform of a cumulative distribution and (ii) a reference link power gain-based framework which exploits the distribution of the fading power gain between the reference transmitter and receiver. The outage probability is spatially averaged over both the fading distribution and the possible locations of the interferers. The boundary effects are accurately accounted for using the probability distribution function of the distance of a random node from the reference receiver. For the case of the node locations modeled by a Binomial point process and Nakagami-$m$ fading channel, we demonstrate the use of the proposed frameworks to evaluate the outage probability at any location inside either a disk or polygon region. The analysis illustrates the location dependent performance in finite wireless networks and highlights the importance of accurately modeling the boundary effects.
\end{abstract}


\ifCLASSOPTIONpeerreview
    \newpage
\fi

\section{Introduction}

\subsection{Motivation}
Outage probability is an important performance metric for wireless networks operating over fading channels~\cite{Haenggi-2012}. It is commonly defined as the probability that the signal-to-interference-plus-noise ratio (SINR) drops below a given threshold. The analysis of the outage probability and interference in wireless networks has received much attention recently~\cite{Weber-2005,Haenggi-2009,Win-2009,Cardieri-2010,Weber-2010,Baccelli-2010,Andrews-2012,Lee-2012}. For the sake of analytical convenience and tractability, all the aforementioned studies and many references therein assumed infinitely large wireless networks and often used a homogeneous Poisson point process (PPP) as the underlying model for the spatial node distribution. A homogeneous PPP is stationary i.e., the node distribution is invariant under translation. This gives rise to location-independent performance, i.e., statistically the network characteristics (such as mean aggregate interference and average outage probability) as seen from a node's perspective are the same for all nodes. Mathematical tools from stochastic geometry have been applied to obtain analytical expressions for the outage probability in infinitely large wireless networks~\cite{Haenggi-2012,Baccelli-2010,Andrews-2012}. The outage analysis in infinite wireless networks has also been extended to wireless networks with the Poisson cluster process~\cite{Ganti-2009,Haenggi-2009,Baccelli-2010}, as well as to coexisting networks sharing the same frequency spectrum~\cite{Kahlon-11,ElSawy-2013,Lee-2012b,Vijayandran-2012}.

In practice, many real-world wireless networks comprise a finite number of nodes distributed at random inside a given \textit{finite} region. The boundary effect of finite networks gives rise to non-stationary location-dependent performance, i.e., the nodes located close to the physical boundaries of the wireless network experience different network characteristics (such as mean aggregate interference and average outage probability) as compared to the nodes located near the center of the network. As a result, the modeling and performance analysis of finite wireless networks requires different approaches as opposed to infinite wireless networks. For example, when a finite number of nodes are independently and uniformly distributed (i.u.d.) inside a finite network, a Binomial point process (BPP), rather than a PPP, provides an accurate model for the spatial node distribution~\cite{Srinivasa-2010,Zubair-2013}. Unlike infinite wireless networks, deriving general results on the outage probability in finite wireless networks is a very difficult task because the outage performance depends strongly on the shape of the network region as well as the location of the reference receiver. In this work, we would like to investigate whether there exist general frameworks that provide easy-to-follow procedures to derive the outage probability at an arbitrary location in an arbitrarily-shaped finite wireless network.

\subsection{Related work}
Since it is difficult to derive general results on the outage probability in arbitrarily-shaped finite wireless networks, most prior works focused on a specific shape (such as disk) and computed the outage probability at a specific location (such as the center of the network region). A few recent studies presented outage characterizations at the center of a BPP network with more general shapes~\cite{Srinivasa-2007,Srinivasa-2010,Chen-2012}. Specifically, the analytical expression for the moment generating function of the aggregate interference seen at the center of a BPP network was presented in~\cite{Srinivasa-2007}. The results were extended for the case of spatial multiplexing with the maximal ratio combining and zero forcing schemes in~\cite{Chen-2012}. Using the probability distribution function (PDF) of the nearest neighbor in a BPP, a lower bound on the outage probability at the center of the wireless network for a simple path loss model was computed in~\cite{Srinivasa-2010}. The exact closed-form outage probability in a class of networks with isotropic node distribution (i.e., the node distribution is invariant under rotation) was derived in~\cite{Tanbourgi-2012}. For finite networks, the results in~\cite{Tanbourgi-2012} can only be applied to very limited cases preserving the isotropic node distribution, such as a disk-shaped network. The outage probability at an arbitrary location of an arbitrarily-shaped finite wireless network was studied in~\cite{Torrieri-2012}. This work focused on deriving closed-form expressions for the \textit{conditional} outage probability, which is conditioned on the locations of all the interfering nodes in the network. For the (unconditional) outage probability averaged over the spatial distribution of nodes, the authors in~\cite{Torrieri-2012} presented the analytical result for the special case of an annular-shaped network with the reference receiver at the center. For other shapes and receiver locations, the authors in~\cite{Torrieri-2012} suggested the use of Monte Carlo simulations to compute the outage probability. Therefore, it is still largely an open research problem to find general frameworks for deriving the outage probability at an arbitrary location of a finite wireless network with an arbitrary shape.

\ifCLASSOPTIONonecolumn
\vspace{-0.3 cm}
\fi

\subsection{Contributions}
In this paper, we focus on the outage probability analysis for a reference transmitter-receiver link in the presence of $M$ interferers and additive white Gaussian noise in a finite wireless network. We present analytical frameworks for computing the outage probability at an arbitrary location in an arbitrarily-shaped finite wireless network. The outage probability is spatially averaged over both the fading distribution and the possible locations of the interferers. The spatial averaging means that the outage probability is not tied to a particular realization of the network and the channel conditions. Specifically, we make the following contributions in the paper:
\begin{itemize}
  \item We propose two general frameworks for the exact calculation of the outage probability in arbitrarily-shaped finite wireless networks in which the reference receiver can be located anywhere.
\begin{itemize}
  \item The first framework, named the moment generating function-based (MGF-based) framework, is based on the numerical inversion of the Laplace transform of the cumulative distribution function (CDF) of an appropriately defined random variable (related to the SINR at the reference receiver). It is inspired from the mathematical techniques developed in~\cite{Abate-1995,Ko-2000} and is valid for any spatial node distribution and any fading channel distribution. To the best of our knowledge, this is the first time that such an approach has been applied in the context of finite wireless networks.

  \item The second framework, named the reference link power gain-based (RLPG-based) framework, exploits the distribution of the fading power gain between the reference transmitter and receiver. It is based on the combination and generalization of the frameworks proposed in~\cite{Srinivasa-2007,Torrieri-2012} and is valid for any spatial node distribution and a general class of fading channel distribution proposed in~\cite{Hunter-2008}.
\end{itemize}

\item In order to demonstrate the use of the proposed frameworks, we consider the case where the nodes are independently and uniformly distributed in the finite wireless network, with node locations modeled by a uniform BPP. The fading channels between all links are assumed to be independently and identically distributed (i.i.d.) according to a Nakagami-$m$ distribution. We use the probability distribution function of the distance of a random node from the reference receiver in uniform BPP networks~\cite{Zubair-2013} to accurately capture the boundary effects in the two frameworks. A summary of the main outage probability results in this paper is shown in Table~\ref{tb:4}.
\begin{table*}[t]
\centering
\caption{Summary of the Main Outage Probability Results.}\label{tb:4}
\begin{tabular}{|c||c|c|c|c|l|} \hline
\multirow{2}{*}{}&\multirow{2}{*}{Equation}& \multirow{2}{*}{\makecell[c]{Node\\distribution}}& \multirow{2}{*}{Region}& \multirow{2}{*}{\makecell[c]{Reference receiver\\location}}& \multirow{2}{*}{Fading channels} \\
&&&&&\\\hline
\multirow{5}{*}{\begin{sideways}MGF-based\end{sideways}}
\multirow{5}{*}{\begin{sideways}Framework\end{sideways}}
& \eqref{final_outage}& any &any& any &any \\
& \eqref{final_outage1}& i.i.d. &any& any &i.i.d. \\
& \eqref{final_gama_outage}&i.i.d. &any& any &i.i.d. Nakagami-$m$ \\
& \eqref{final_gama_outage} \&  \eqref{mgf_disk_expectation}& i.u.d. &disk& any &i.i.d. Nakagami-$m$ \\
& \eqref{final_gama_outage} \&  \eqref{mgf_center_poly}& i.u.d. &polygon& center &i.i.d. Nakagami-$m$ \\\hline
\multirow{5}{*}{\begin{sideways}RLPG-based\end{sideways}}
\multirow{5}{*}{\begin{sideways}Framework\end{sideways}}
& \eqref{ana_outage_form1}& any &any& any &Reference link: \eqref{ana_general_g}~\cite{Hunter-2008} \& Interference links: any  \\
& \eqref{ana_outage_form}& i.i.d. & any&any &Reference link: \eqref{ana_general_g}~\cite{Hunter-2008} \& Interference links: i.i.d.  \\
& \eqref{ana_gamma_outage}&i.i.d. & any &any &i.i.d. Nakagami-$m$ with integer $m_0$ for the reference link \\
& \eqref{ana_gamma_outage} \&  \eqref{e_disk}& i.u.d. &disk& any &i.i.d. Nakagami-$m$ with integer $m_0$ for the reference link \\
& \eqref{ana_gamma_outage} \&  \eqref{ana_center_e}& i.u.d. &polygon& center &i.i.d. Nakagami-$m$ with integer $m_0$ for the reference link \\
\hline

\end{tabular}
\end{table*}

\begin{itemize}
  \item We demonstrate the use of the two frameworks in evaluating the outage probability for the important case of a reference receiver located anywhere in a
  disk region. We show that the known outage probability results in the literature for the disk region arise as special cases in our two frameworks.

  \item We further consider the case of an arbitrary located reference receiver in a convex polygon region and present an algorithm for accurately computing the outage probability. This fundamentally extends the prior work dealing with the outage probability analysis in finite wireless networks.

  \item We show that the impact of boundary effects in finite wireless networks is location dependent and is enhanced by an increase in the $m_0 = m$ value for Nakagami-$m$ fading channels or an increase in the path-loss exponent. We also show that due to the boundary effects, the outage probability using PPP model does not provide any meaningful bounds for the outage probability in an arbitrarily-shaped finite region. This highlights the importance of the proposed frameworks, which allow accurate outage probability computation for arbitrarily-shaped finite wireless networks.
\end{itemize}
\end{itemize}

\subsection{Notation and Paper Organization}
The following notation is used in this paper. $\Pr(\cdot)$ indicates the probability measure and $\mathrm{E}\{\cdot\}$ denotes the expectation operator. $\mathbf{i}$ is the imaginary number and $\mathrm{Re}\{\cdot\}$ denotes the real part of a complex-valued number. $|\mathcal{A}|$ denotes the area of an arbitrarily-shaped finite wireless network region $\mathcal{A}$. $\Gamma(x)=\int_0^{\infty}t^{x-1}\exp(-t) dt$ and $\Gamma(a,x)=\int_a^{\infty}t^{x-1}\exp(-t) dt$ are the complete gamma function and incomplete upper gamma functions, respectively~\cite{gradshteyn2007}. $_{2}F_{1}\left(\cdot,\cdot,\cdot,\cdot\right)$ is the Gaussian or ordinary hypergeometric function~\cite{gradshteyn2007}. $f_Z(z)$ and $F_Z(z)$ denotes the probability distribution function and the cumulative distribution function of a random variable $Z$. $\mathcal{L}_Z(s)=\mathcal{L}_{f_Z(z)}(s)=\mathrm{E}\{\exp(-sZ)\}$ denotes the Laplace transform of the probability distribution function $f_Z(z)$. $\mathcal{L}_{F_Z(z)}(s) = \mathcal{L}_{f_Z(z)}(s)/s = \mathcal{L}_Z(s)/s$  denotes the Laplace transform of the cumulative distribution function of the random variable $Z$. Table~\ref{tb:2} summarizes the main mathematical symbols and random variables (RVs) used in this paper.

The remainder of the paper is organised as follows: The system model and assumptions are presented in Section~\ref{problemform}. The proposed general frameworks are described in detail in Section~\ref{framework}. For the case of the node locations modeled by a Binomial point process and i.i.d. Nakagami-$m$ fading channels, the evaluation of the proposed frameworks for arbitrarily-shaped finite wireless networks is detailed in Section~\ref{gam}. The outage probability analysis for the case of a reference receiver located anywhere in a disk and a convex polygon is presented in Section~\ref{disk} and Section~\ref{other}, respectively. The derived results are used to study the outage probability in Section~\ref{result}. Finally, the paper is summarized in Section~\ref{conclusion}.
%
\ifCLASSOPTIONonecolumn
\begin{table}[t]
\centering
\caption{Summary of the Main Mathematical Symbols.}\label{tb:2}
\begin{tabular}{|c|c|l|c|c|l|} \hline
 & Symbol& Description & & Symbol& Description \\ \hline %
\multirow{16}{*}{\begin{sideways}System parameters\end{sideways}}
& $M$ & Number of interfering nodes & \multirow{5}{*}{\begin{sideways}RVs\end{sideways}} &$X_i$  & $i$th interfering node and its location \\
&$\area$ & Arbitrarily-shaped finite region & &$R_i$ & Euclidean distance between $X_i$ and $Y_0$\\
&$\mathcal{R}$ & Radius of disk or circum-radius of a polygon  & &$G_0$ & Power gain due to fading for the reference link \\
&$X_0$ & Reference transmitter and its location & &$G_i$ & Power gain due to fading for the $i$th interference link \\
&$Y_0$ & Reference receiver and its location & &$I$ & Aggregate interference \\
&$r_0$ & Euclidean distance between $X_0$ and $Y_0$ & &$\gamma$ & Signal-to-interference-plus-noise ratio (SINR) \\
&$P_0$ &Transmit power for reference transmitter $X_0$ & & & \\
&$P_i$ & Transmit power for interferer $X_i$ & & & \\
&$\alpha$ & Path-loss exponent & & & \\
&$m_0$ & Nakagami-$m$ fading parameter for reference link & & & \\
&$m$ & Nakagami-$m$ fading parameter for interference links & & & \\
&$N$ & Additive white Gaussian noise (AWGN) power & & & \\
&$\rho_0$ & Signal-to-noise ratio & & & \\
&$\beta$ & SINR threshold & & & \\
&$\epsilon$ & Outage probability & & & \\ \hline
\end{tabular}
\end{table}
\else
\begin{table}[t]
\centering
\caption{Summary of the Main Mathematical Symbols.}\label{tb:2}
\begin{tabular}{|c||c|l|} \hline
 & Symbol& Description \\ \hline %
\multirow{16}{*}{\begin{sideways}System parameters\end{sideways}}
& $M$ & Number of interfering nodes \\
&$\area$ & Arbitrarily-shaped finite region \\
&$\radius$ & Radius of disk or circum-radius of a polygon\\
&$X_0$ & Reference transmitter and its location \\
&$Y_0$ & Reference receiver and its location \\
&$r_0$ & Euclidean distance between $X_0$ and $Y_0$ \\
&$P_0$ &Transmit power for reference transmitter $X_0$ \\
&$P_i$ & Transmit power for interferer $X_i$ \\
&$\alpha$ & Path-loss exponent \\
&$m_0$ & Nakagami-$m$ fading parameter for reference link \\
&$m$ & Nakagami-$m$ fading parameter for interference links \\
&$N$ & Additive white Gaussian noise (AWGN) power \\
&$\rho_0$ & Signal-to-noise ratio \\
&$\beta$ & SINR threshold \\
&$\epsilon$ & Outage probability \\\hline
\multirow{5}{*}{\begin{sideways}RVs\end{sideways}}
&$X_i$  & $i$th interfering node and its location \\
&$R_i$ & Euclidean distance between $X_i$ and $Y_0$ \\
&$G_0$ & Power gain due to fading for the reference link \\
&$G_i$ & Power gain due to fading for the $i$th interference link \\
&$I$ & Aggregate interference \\
&$\gamma$ & Signal-to-interference-plus-noise ratio (SINR) \\ \hline
\end{tabular}
\end{table}
\fi


\section{Problem Formulation and System Model Assumptions}\label{problemform}
Consider a wireless network with $M+2$ nodes which are located inside an arbitrarily-shaped finite region $\area \subset \mathbb{R}^2$, where $\mathbb{R}^2$ denotes the two-dimensional Euclidean domain. The $M+2$ nodes consist of a reference transmitter $X_0$, a reference receiver $Y_0$ and $M$ interfering nodes. The $M$ interfering nodes are distributed at random within the region $\area$. Throughout the paper, we refer to $X_i$ $(i=1,2,\hdots M)$ as both the random location as well as the $i$th interfering node itself. The reference receiver $Y_0$ is not restricted to be located at the center of the finite region but can be located \textit{anywhere} inside the region $\area$. The reference transmitter $X_0$ is assumed to be placed at a given distance $r_0$ from $Y_0$. Let $R_i$ $(i=1,2,\hdots M)$ denote the Euclidean distance between the $i$th interferer $X_i$ and the reference receiver $Y_0$.

We focus on the performance of the reference link comprising the reference transmitter $X_0$ and the reference receiver $Y_0$, in the presence of $M$ interfering nodes and noise. We consider a path-loss plus block-fading channel model. Let $G_0$ represent the instantaneous power gain due to fading only for the reference link and $G_i$ represent the instantaneous power gain due to fading only between $X_i$ and $Y_0$. The path-loss function can be expressed as
\begin{align}\label{path_loss_f}
l(r)=r^{-\alpha},
\end{align}
where $r$ denotes the distance and $\alpha$ is the path-loss exponent, which typically lies in the range $2\leq\alpha\leq6$~\cite{Marvin2005}. Note that the path loss model in~\eqref{path_loss_f} is unbounded and has a singularity as $r \rightarrow 0$. The singularity can be avoided by using a bounded path loss model~\cite{Inaltekin-2009}. Because we consider the network from an outage perspective, the effect of the singularity in the bounded path loss model is in fact negligible~\cite{Weber-2007}, as long as the SINR threshold (defined in~\eqref{outage_define}) is not too small. Thus, for simplicity, we can adopt the unbounded path-loss model for the purpose of outage probability computation.

Let $P_0$ and $P_i$ denote the transmit powers for $X_0$ and $X_i$, respectively. The aggregate interference power at the reference receiver $Y_0$ is then given as
\begin{align}\label{aggregat_I}
 I=\sum_{i=1}^{M}P_iG_iR_i^{-\alpha}.
\end{align}

The instantaneous signal-to-interference-plus-noise ratio~$\gamma$ at the reference receiver $Y_0$ is given by
\begin{align}\label{sinr_define}
\gamma =\frac{P_0G_0r_0^{-\alpha}}{N+I} =\frac{G_0}{\frac{1}{\rho_0}+\frac{I}{P_0r_0^{-\alpha}}},
\end{align}

\noindent where $\rho_0=(P_0r_0^{-\alpha})/N$ is defined as the average signal-to-noise ratio (SNR) and $N$ is the additive white Gaussian noise (AWGN) power.

We characterize the performance of the reference link in the presence of the aggregate interference and AWGN by using the outage probability. An outage is said to occur when the SINR $\gamma$ falls below a given SINR threshold $\beta$, i.e.,
\begin{align}\label{outage_define}
\epsilon =\Pr(\gamma<\beta).
\end{align}

\noindent We are interested in obtaining the average outage probability in the arbitrarily-shaped finite wireless network after un-conditioning with respect to the spatial node distribution and the fading distribution. This is addressed in the next section.


\section{Proposed Frameworks}\label{framework}

In this section, we propose two analytical frameworks to compute the outage probability in arbitrarily-shaped finite wireless network. The first framework, named the moment generating function-based (MGF-based) framework, is inspired from~\cite{Abate-1995,Ko-2000}. The basic principle of this framework is the accurate numerical inversion of the Laplace transform of the cumulative distribution function for an appropriately defined random variable~\cite{Abate-1995}. The second framework, named the reference link power gain-based (RLPG-based) framework, is based on the combination and generalization of the frameworks proposed in~\cite{Srinivasa-2007,Torrieri-2012}. The basic principle of this second framework is to find the cumulative distribution function of the reference link's fading power gain, which can then be used to find the outage probability. These frameworks are discussed in detail in the following subsections.

\subsection{Moment Generating Function-based (MGF-based) Framework}
In this framework, it is necessary to define a suitable random variable. Substituting~\eqref{sinr_define} into~\eqref{outage_define} and rearranging,
we have
\begin{align}\label{LT_define_outage}
\epsilon=\Pr\left(\frac{1}{\rho_0G_0}+\frac{I}{P_0r_0^{-\alpha}G_0}>\beta^{-1}\right).
\end{align}

\noindent We define a random variable $Z$ as
\begin{align}\label{z_define}
Z=\frac{1}{\rho_0G_0}+\frac{I}{P_0r_0^{-\alpha}G_0}.
\end{align}

\noindent Hence,~\eqref{LT_define_outage} can be re-written as
\begin{align}\label{LT_new_outage}
\epsilon =\Pr(Z>\beta^{-1})=1-F_Z(\beta^{-1}).
\end{align}

In general, it is not possible to obtain a closed-form solution for $F_Z(\beta^{-1})$. Hence, we use numerical inversion of Laplace transform to find
$F_Z(\beta^{-1})$. The CDF of a random variable $Z$ is related to the Laplace transform of $F_Z(z)$ as $F_Z(z)=\frac{1}{2\pi\jj}\int_{a-\jj\infty}^{a+\jj\infty}\mathcal{L}_{F_Z(z)}(s)\exp(s z)ds$. Using the trapezoid rule, the above integral can be discretized to get a series and then we can truncate the infinite series to get a finite sum via the Euler summation~\cite{OCinneide-1997}. Finally, since~$\mathcal{L}_{F_Z(z)}(s) = \mathcal{L}_Z(s)/s$,~\eqref{LT_new_outage} can be approximated by
\ifCLASSOPTIONonecolumn
\begin{align}\label{LT_main}
\epsilon&=1-\frac{2^{-B}\exp(\frac{A}{2})}{\beta^{-1}}\sum_{b=0}^{B}\binom{B}{b} \sum_{c=0}^{C+b}\frac{(-1)^c}{D_c}\mathrm{Re} \left\{\frac{\mathcal{L}_Z\left(s\right)}{s} \right\},
\end{align}
\else
\begin{align}\label{LT_main}
\epsilon&=1-\frac{2^{-B}\exp(\frac{A}{2})}{\beta^{-1}}\sum_{b=0}^{B}\binom{B}{b} \sum_{c=0}^{C+b}\frac{(-1)^c}{D_c}\mathrm{Re} \left\{\frac{\mathcal{L}_Z\left(s\right)}{s} \right\},
\end{align}
\fi

\noindent where $D_c= 2$ (if $c=0$) and $D_c=1$ (if $c=1,2,\hdots$), $s=(A+\mathbf{i}2\pi c)/(2\beta^{-1})$ and $\mathrm{Re}\{\cdot\}$ denotes the real part. The three parameters $A$, $B$ and $C$ control the estimation error. The selection of the values for $A$, $B$ and $C$ for accurate numerical inversion will be discussed later in Section~\ref{compute}.

Using the definition of the Laplace transform of the probability distribution of a random variable, we can express $\mathcal{L}_Z(s)$ as
\ifCLASSOPTIONonecolumn
\begin{align}\label{LT_z_lt}
\mathcal{L}_Z(s)=\mathrm{E}_{G_0,I}\left\{\exp\left(-s\left(\frac{1}{\rho_0G_0}+\frac{I}{P_0r_0^{-\alpha}G_0}\right)\right)\right\}=\mathrm{E}_{G_0,G_i,R_i}\left\{\exp\left(-\frac{s}{\rho_0G_0}\right)\prod_{i=1}^{M}\exp\left(\frac{-sP_iG_iR_i^{-\alpha}}{P_0r_0^{-\alpha}G_0}\right)\right\},
\end{align}
\else
\begin{small}
\begin{align}\label{LT_z_lt}
\mathcal{L}_Z(s)&=\mathrm{E}_{G_0,I}\left\{\exp\left(-s\left(\frac{1}{\rho_0G_0}+\frac{I}{P_0r_0^{-\alpha}G_0}\right)\right)\right\} \nonumber \\
&=\mathrm{E}_{G_0,G_i,R_i}\left\{\exp\left(-\frac{s}{\rho_0G_0}\right)\prod_{i=1}^{M}\exp\left(\frac{-sP_iG_iR_i^{-\alpha}}{P_0r_0^{-\alpha}G_0}\right)\right\},
\end{align}
\end{small}
\fi

\noindent where $\mathrm{E}_I\{\cdot\}$ denotes the expectation with respect to the aggregate interference
and $\mathrm{E}_{G_i,R_i}\{\cdot\}$ represents the expectation with respect to $G_i$ and $R_i$. Combining~\eqref{LT_main} and~\eqref{LT_z_lt}, we have the general outage probability expression resulting from the MGF-based framework as
%
\ifCLASSOPTIONonecolumn
\begin{align}\label{final_outage}
\epsilon = 1- \frac{2^{-B}\exp(\frac{A}{2})}{\beta^{-1}}\sum_{b=0}^{B}\binom{B}{b}\sum_{c=0}^{C+b}\frac{(-1)^c}{D_c}\textrm{Re} \left\{\frac{\mathrm{E}_{G_i,R_i}\left\{\exp\left(-\frac{s}{\rho_0G_0}\right)\prod\limits_{i=1}\limits^{M}\exp\left(\frac{-sP_iG_iR_i^{-\alpha}}{P_0r_0^{-\alpha}G_0}\right)\right\}}{s} \right\}.
\end{align}
\else
\begin{align}\label{final_outage}
\epsilon =& 1- \frac{2^{-B}\exp(\frac{A}{2})}{\beta^{-1}}\sum_{b=0}^{B}\binom{B}{b}\sum_{c=0}^{C+b}\frac{(-1)^c}{D_c}\times\\ \nonumber
&\textrm{Re} \left\{\frac{\mathrm{E}_{G_i,R_i}\left\{\exp\left(-\frac{s}{\rho_0G_0}\right)\prod\limits_{i=1}\limits^{M}\exp\left(\frac{-sP_iG_iR_i^{-\alpha}}{P_0r_0^{-\alpha}G_0}\right)\right\}}{s} \right\}.
\end{align}
\fi

\subsection{Reference Link Power Gain-based (RLPG-based) Framework}
As highlighted earlier, this framework relies on the cumulative distribution function $F_{G_0}(g_0)$ for the reference link's fading power gain $G_0$. We adopt the following general model\footnote{We note that it may be possible to consider other general classes of fading channels~\cite{yacoub-2007,Ahmadi-2009,Yilmaz-2010}, but this is outside the scope of this paper.} for the CDF~\cite{Hunter-2008}
\begin{equation}\label{ana_general_g}
F_{G_0}(g_0)=1-\sum_{n\in \mathcal{N}}\exp(-ng_0)\sum_{k\in \mathcal{K}}a_{nk}g_0^{k},
\end{equation}

\noindent where the finite sets $\mathcal{N},\mathcal{K}\subset \mathbb{N}$ (where $\mathbb{N}$ is the set of natural number) and $a_{nk}$ are the coefficients.

With the proper choice of $a_{nk}$, $\mathcal{N}$ and $\mathcal{K}$, $F_{G_0}(g_0)$ can represent different types of distributions for the power gain due to fading~\cite{Hunter-2008}. For example, when $\mathcal{N}=\{1\}$, $\mathcal{K}=\{0\}$ and $a_{nk}=1$, then $F_{G_0}(g_0)=1-\exp(-g_0)$ and consequently $\fgg$ reduces to the Exponential distribution, which corresponds to the reference link undergoing Rayleigh fading~\cite{Kahlon-2012}. Furthermore, if $\mathcal{N}=\{m_0\}$, $\mathcal{K}=\{0,...,m_0-1\}$ and $a_{nk}=\frac{m_0^k}{k!}$, then $F_{G_0}(g_0)=1-\exp(-m_0 g_0)\sum_{k=0}^{m_0-1}m_0^k g_0^k/k!$ and consequently $\fgg$ reduces to the Gamma distribution~\cite{robert2004} which corresponds to the reference link experiencing Nakagami-$m$ fading.

We proceed by re-writing the outage probability in~\eqref{outage_define} as
\ifCLASSOPTIONonecolumn
\begin{align}\label{ana_define_outage}
\epsilon = \mathrm{E}_I\left\{\Pr\left(\left.\frac{G_0}{\frac{1}{\rho_0}+\frac{I}{P_0r_0^{-\alpha}}}<\beta \right| I \right)\right\} =\mathrm{E}_I\left\{\Pr\left(\left.G_0<\beta\left(\frac{1}{\rho_0}+\frac{I}{P_0r_0^{-\alpha}}\right)\right|I \right)\right\}.
\end{align}
\else
\begin{align}\label{ana_define_outage}
\epsilon &= \mathrm{E}_I\left\{\Pr\left(\left.\frac{G_0}{\frac{1}{\rho_0}+\frac{I}{P_0r_0^{-\alpha}}}<\beta \right| I \right)\right\} \nonumber\\
&=\mathrm{E}_I\left\{\Pr\left(\left.G_0<\beta\left(\frac{1}{\rho_0}+\frac{I}{P_0r_0^{-\alpha}}\right)\right|I \right)\right\}.
\end{align}
\fi

Using $F_{G_0}(g_0)$ shown in \eqref{ana_general_g}, \eqref{ana_define_outage} can be written as
\ifCLASSOPTIONonecolumn
\begin{align}\label{ana_outage}
\epsilon =& \mathrm{E}_I\left\{F_{G_0}\left(\A+\B I\right) \right\}\nonumber\\
=& 1-\sum_{n\in \mathcal{N}}\exp\left(-n\A\right)\sum_{k\in \mathcal{K}}a_{nk}\mathrm{E}_I\left\{\exp\left(-n\B I\right)\left(\A+\B I\right)^{k}\right\},
\end{align}
\else
\begin{align}\label{ana_outage}
\epsilon =& \mathrm{E}_I\left\{F_{G_0}\left(\A+\B I\right) \right\}\nonumber\\
=& 1-\sum_{n\in \mathcal{N}}\exp\left(-n\A\right)\sum_{k\in \mathcal{K}}a_{nk}\times\nonumber\\
&\mathrm{E}_I\left\{\exp\left(-n\B I\right)\left(\A+\B I\right)^{k}\right\},
\end{align}
\fi
\noindent where the aggregate interference $I$ depends on the fading power gain distribution and the distance distribution. Hence, \eqref{ana_outage} can be further expanded in terms of $G_i$ and $R_i$. First, we focus on the expansion of the term $\exp\left(-n\B I\right)\left(\A+\B I\right)^{k}$. Based on the binomial theorem~\cite{Abramowitz1972} and using~\eqref{aggregat_I}, we have
\ifCLASSOPTIONonecolumn
\begin{align}\label{ana_binomial}
&\exp\left(-n\B I\right)\left(\A+\B I\right)^{k}  \nonumber\\
=&\exp\left(-n\B\sum_{i=1}^{M}P_iG_iR_i^{-\alpha}\right)\sum_{j=0}^{k}\binom{k}{j}\left(\A\right)^{k-j}\left(\B\right)^j\left(\sum_{i=1}^{M}P_iG_iR_i^{-\alpha}\right)^j.
\end{align}
\else
\begin{align}\label{ana_binomial}
&\exp\left(-n\B I\right)\left(\A+\B I\right)^{k}  \nonumber\\
=&\exp\left(-n\B\sum_{i=1}^{M}P_iG_iR_i^{-\alpha}\right)\sum_{j=0}^{k}\binom{k}{j}\times\nonumber\\
&\left(\A\right)^{k-j}\left(\B\right)^j\left(\sum_{i=1}^{M}P_iG_iR_i^{-\alpha}\right)^j.
\end{align}
\fi

\noindent Then, following the multinomial theorem~\cite{multiserires2013}, we can further expand the term $\left(\sum_{i=1}^{M}P_iG_iR_i^{-\alpha}\right)^j$ into
\ifCLASSOPTIONonecolumn
\begin{align}\label{ana_multinomial}
\left(\sum_{i=1}^{M}P_iG_iR_i^{-\alpha}\right)^j=\sum_{t_1+t_2+...+t_M=j}\binom{j}{t_1,t_2,...,t_M}\prod_{i=1}^{M}\left(P_iG_iR_i^{-\alpha}\right)^{t_i},
\end{align}
\else
\begin{small}
\begin{align}\label{ana_multinomial}
&\left(\sum_{i=1}^{M}P_iG_iR_i^{-\alpha}\right)^j\nonumber\\
=&\sum_{t_1+t_2+...+t_M=j}\binom{j}{t_1,t_2,...,t_M}\prod_{i=1}^{M}\left(P_iG_iR_i^{-\alpha}\right)^{t_i},
\end{align}
\end{small}
\fi

\noindent where $t_i$ $(i=1,2,...M)$ is a non-negative integer and the  multinomial coefficient $\binom{j}{t_1,t_2,...,t_M}=\frac{j!}{t_1!t_2!...t_M!}$.

Combining \eqref{ana_binomial} and \eqref{ana_multinomial} and substituting back into \eqref{ana_outage}, we have the general outage probability expression resulting from the RLPG-based framework as
\ifCLASSOPTIONonecolumn
\begin{align}\label{ana_outage_form1}
\epsilon =& 1-\sum_{n\in \mathcal{N}}\exp\left(-n\A\right)\sum_{k\in \mathcal{K}}a_{nk}\sum_{j=0}^{k}\binom{k}{j}\left(\A\right)^{k-j}\left(\B\right)^j \sum_{t_1+t_2...+t_M=j}\binom{j}{t_1,t_2,...,t_M}\times \nonumber\\
&\mathrm{E}_{G_i,R_i}\left\{\exp\left(-n\B\sum_{i=1}^{M}P_iG_iR_i^{-\alpha}\right)\prod_{i=1}^{M}\left(P_iG_iR_i^{-\alpha}\right)^{t_i}  \right\}.
\end{align}
\else
\begin{small}
\begin{align}\label{ana_outage_form1}
\epsilon =& 1-\sum_{n\in \mathcal{N}}\exp\left(-n\A\right)\sum_{k\in \mathcal{K}}a_{nk}\sum_{j=0}^{k}\binom{k}{j}\times\nonumber\\
&\left(\A\right)^{k-j}\left(\B\right)^j \sum_{t_1+t_2...+t_M=j}\binom{j}{t_1,t_2,...,t_M}\times \nonumber\\
&\mathrm{E}_{G_i,R_i}\left\{\exp\left(-n\B\sum_{i=1}^{M}P_iG_iR_i^{-\alpha}\right)\prod_{i=1}^{M}\left(P_iG_iR_i^{-\alpha}\right)^{t_i}  \right\}.
\end{align}
\end{small}
\fi

\begin{remark}\label{remark1}
The two general formulations in~\eqref{final_outage} and~\eqref{ana_outage_form1} are valid for any spatial node distribution with a fixed number of nodes in an arbitrarily-shaped finite wireless network and any location of the reference receiver inside the arbitrarily-shaped finite region.
The MGF-based framework in~\eqref{final_outage} is valid for any fading channel distribution. The RLPG-based framework in~\eqref{ana_outage_form1} is valid for a general class of fading channel distribution defined in~\eqref{ana_general_g}. In general, the two formulations cannot be expressed in closed-form. The evaluation of the outage probability in~\eqref{final_outage} and~\eqref{ana_outage_form1} requires the knowledge of the joint probability distribution function of the distance $R_i$ and the fading power gain $G_i$. For an arbitrary location of a reference point inside an arbitrarily-shaped convex region, the joint probability distribution function can be a complex piece-wise function, which does not allow the outage probability to be computed in a closed-form. However, it must be noted that~\eqref{final_outage} and~\eqref{ana_outage_form1} give analytical expressions of the outage probability for arbitrarily-shaped finite wireless networks in a general setting, which has not been demonstrated in the literature to date.
\end{remark}

In the next section, in order to demonstrate the evaluation and use of the proposed frameworks, we consider the case of that the nodes are distributed at random inside an arbitrarily-shaped finite wireless network according to a BPP and all fading channels are i.i.d. Nakagami-$m$ fading channels.

\section{Outage probability for BPP and Nakagami-$m$ Fading Channels}\label{gam}

In this section, we consider the scenario where:
\begin{enumerate}[{A}1.]
\item The $M$ interfering nodes are independently and identically (i.i.d.) distributed at random inside an arbitrarily-shaped finite wireless network, i.e. the nodes are distributed at random according to a general BPP~\cite[Definition 2.12]{Haenggi-2012}.

\item The fading channels are independently and identically distributed (i.i.d.).

\item The transmit powers $P_0$ and $P_i$ are normalized to unity.

\item The fading channels follow a Nakagami-$m$ distribution. The Nakagami-$m$ distribution is widely used in the literature to model the distribution of the signal envelopes in various fading environments, such as the land-mobile and indoor-mobile multipath propagation environments~\cite{Marvin2005}. The parameter $m$, which lies in the range $1/2$ to $\infty$, describes the severity of the fading channel. Note that it is not necessary for $m$ to be an integer number only. When $m\leq 1$, the Nakagami-$m$ distribution provides a close approximation to the Nakagami-$q$ (Hoyt) distribution with parameter mapping $m=((1+q^2)^2)/(2(1+2q^4))$. Additionally, when $m > 1$, the Nakagami-$m$ distribution closely approximates the Nakagami-$n$ (Rice) distribution with parameter mapping $m=((1+n^2)^2)/(1+2n^2)$~\cite{Marvin2005}. It is well known that $m=\infty$ corresponds to the no-fading case, $m=1$ represents the special case of Rayleigh fading and $m=1/2$ represents that unilateral Gauss distribution~\cite{Marvin2005}, which corresponds to the most severe Nakagami-$m$ fading.

\item The nodes are distributed at random according to a \textit{uniform} BPP~\cite[Definition 2.11]{Haenggi-2012}. This means that the nodes are independently and uniformly distributed (i.u.d.) inside the arbitrarily-shaped finite wireless network.

\end{enumerate}

As a consequence of assumptions A1 and A2, the joint PDF of the distance $R_i$ and the fading power gain $G_i$ can be decomposed into the individual PDFs, which are denoted as $f_{R_i}(r_i)$ and $f_{G_i}(g_i)$, respectively. Due to assumption A1, the distribution of $R_i$ is the same for all $i$. Similarly, due to assumption A2, the distribution of $G_i$ is the same for all $i$ as well. Thus, we can drop the index $i$ in $R_i$, $G_i$, $f_{R_i}(r_i)$ and $f_{G_i}(g_i)$ and let $f_{R_i}(r_i)=f_{R}(r)$ and $f_{G_i}(g_i)=f_{G}(g)$.

Let $f_{G_0}(g_0)$ denotes the PDF of fading power gain for the reference link. From assumption A4, since the fading coefficients for both the reference link and the interference links are modeled using a Nakagami-$m$ distribution, i.e., $\fgg=\fg$, the distribution for the fading power gains $G_0$ and $G$ can be modeled as a Gamma distribution with the following PDF~\cite{robert2004}
\begin{align}\label{g_i_distri}
\fg=\frac{g^{m-1}m^m}{\Gamma(m)}\exp(-m g).
\end{align}

\noindent Note that we will represent the Nakagami-$m$ fading parameter for the reference and interfering links as $m_0$ and $m$, respectively. Even though the fading power gain distributions are identical, we will still use $m_0$ and $m$ to distinguish the reference link from the interference link for the sake of analytical convenience.

\ifCLASSOPTIONonecolumn
\vspace{-0.35 cm}
\fi

\subsection{MGF-based Framework}

\ifCLASSOPTIONonecolumn
Using assumptions A1$-$A3, i.e., i.i.d. random nodes and i.i.d. fading channels, the general outage probability expression in~\eqref{final_outage} for the MGF-based framework can be simplified to
\begin{align}\label{final_outage1}
\epsilon = 1- \frac{2^{-B}\exp(\frac{A}{2})}{\beta^{-1}}\sum_{b=0}^{B}\binom{B}{b}\sum_{c=0}^{C+b}\frac{(-1)^c}{D_c}\textrm{Re} \left\{\frac{\mathrm{E}_{G_0}\left\{\exp\left(-\frac{s}{\rho_0G_0}\right)\left(\mathrm{E}_{G,R}\left\{\exp\left(\frac{-sGR^{-\alpha}}{r_0^{-\alpha}G_0}\right)\right\}\right)^M\right\}}{s} \right\}.
\end{align}
\else
Using assumptions A1$-$A3, i.e., i.i.d. random nodes and i.i.d. fading channels, the general outage probability expression in~\eqref{final_outage} for the MGF-based framework can be simplified to~\eqref{final_outage1}, shown at the top of the next page.
\begin{figure*}[!t]
\normalsize
\begin{align}\label{final_outage1}
\epsilon = 1- \frac{2^{-B}\exp(\frac{A}{2})}{\beta^{-1}}\sum_{b=0}^{B}\binom{B}{b}\sum_{c=0}^{C+b}\frac{(-1)^c}{D_c}\textrm{Re} \left\{\frac{\mathrm{E}_{G_0}\left\{\exp\left(-\frac{s}{\rho_0G_0}\right)\left(\mathrm{E}_{G,R}\left\{\exp\left(\frac{-sGR^{-\alpha}}{r_0^{-\alpha}G_0}\right)\right\}\right)^M\right\}}{s} \right\}.
\end{align}
\hrulefill
\vspace*{4pt}
\end{figure*}
\fi
\ifCLASSOPTIONonecolumn
Using assumption A4 and substituting~\eqref{g_i_distri} in~\eqref{final_outage1}, the outage probability in~\eqref{final_outage1} can be expressed as
\begin{align}\label{final_gama_outage}
\epsilon =& 1- \frac{2^{-B}\exp(\frac{A}{2})}{\beta^{-1}}\sum_{b=0}^{B}\binom{B}{b}\sum_{c=0}^{C+b}\frac{(-1)^c}{D_c}\times \nonumber\\
&\textrm{Re} \left\{\frac{\int_{0}^{\infty}  \exp\left(-\frac{s}{\rho_0g_0}\right)\left(\mathrm{E}_{G,R}\left\{\exp\left(\frac{-sGR^{-\alpha}}{r_0^{-\alpha}G_0}\right)\right\}\right)^M \frac{g_0^{m_0-1}m_0^{m_0}}{\Gamma(m_0)}\exp(-m_0 g_0) dg_0}{s} \right\},
\end{align}

\noindent where the expectation in~\eqref{final_gama_outage} can be expressed as
\begin{align}\label{LT_gamma_z}
\mathrm{E}_{G,R}\left\{\exp\left(\frac{-sGR^{-\alpha}}{r_0^{-\alpha}G_0}\right)\right\}=\int_0^{r_{\textrm{max}}} m^m\left(m+\frac{r^{-\alpha}r_0^{\alpha}s}{g_0}\right)^{-m} \fr dr,
\end{align}

\noindent where $r_{\textrm{max}}$ denotes the maximum range of the random variable $R$, which depends on the arbitrarily-shaped finite region $\mathcal{A}$ and the location of the reference node. Note that $m_0$ and $m$ in~\eqref{final_gama_outage} can take any values (whether integer or non-integer).
\else
Using assumption A4 and substituting~\eqref{g_i_distri} in~\eqref{final_outage1}, the outage probability in~\eqref{final_outage1} can be expressed as~\eqref{final_gama_outage}, shown at the top of the next page, where the expectation in~\eqref{final_gama_outage} can be expressed as
\begin{figure*}[!t]
\normalsize
\begin{small}
\begin{align}\label{final_gama_outage}
\epsilon = 1- \frac{2^{-B}\exp(\frac{A}{2})}{\beta^{-1}}\sum_{b=0}^{B}\binom{B}{b}\sum_{c=0}^{C+b}\frac{(-1)^c}{D_c}\textrm{Re} \left\{\frac{\int_{0}^{\infty} \exp\left(-\frac{s}{\rho_0g_0}\right)\left(\mathrm{E}_{G,R}\left\{\exp\left(\frac{-sGR^{-\alpha}}{r_0^{-\alpha}G_0}\right)\right\}\right)^M \frac{g_0^{m_0-1}m_0^{m_0}}{\Gamma(m_0)}\exp(-m_0 g_0) dg_0}{s} \right\}.
\end{align}
\end{small}
\hrulefill
\vspace*{4pt}
\end{figure*}
%
\begin{align}\label{LT_gamma_z}
&\mathrm{E}_{G,R}\left\{\exp\left(\frac{-sGR^{-\alpha}}{r_0^{-\alpha}G_0}\right)\right\}= \nonumber\\
&\int_0^{r_{\textrm{max}}} m^m\left(m+\frac{r^{-\alpha}r_0^{\alpha}s}{g_0}\right)^{-m} \fr dr,
\end{align}

\noindent where $r_{\textrm{max}}$ denotes the maximum range of the random variable $R$, which depends on the arbitrarily-shaped finite region $\mathcal{A}$ and the location of the reference node. Note that $m_0$ and $m$ in~\eqref{final_gama_outage} can take any values (integer or non-integer).
\fi

\subsection{RLPG-based Framework}
\ifCLASSOPTIONonecolumn
Using assumptions A1$-$A3, i.e., i.i.d. random nodes and i.i.d. fading channels, the general outage probability expression in~\eqref{ana_outage_form1} for the RLPG-based framework can be simplified to
\begin{align}\label{ana_outage_form}
\epsilon =1-\sum_{n\in \mathcal{N}}\exp\left(-n\A\right)\sum_{k\in \mathcal{K}}a_{nk}\sum_{j=0}^{k}\binom{k}{j}\left(\A\right)^{k-j}\left(\Bnew\right)^j\sum_{t_1+t_2...+t_M=j}\binom{j}{t_1,t_2,...,t_M}\prod_{i=1}^{M}\mathrm{E}_{G,R}\left\{\E\right\},
\end{align}

\noindent where $\E=\exp\left(-n\Bnew GR^{-\alpha}\right)\left(GR^{-\alpha}\right)^{t_i}$ is defined for analytical convenience.
\else
Using assumptions A1$-$A3, i.e., i.i.d. random nodes and i.i.d. fading channels, the general outage probability expression in~\eqref{ana_outage_form1} for the RLPG-based framework can be simplified to~\eqref{ana_outage_form}, shown at the top of the next page, where $\E=\exp\left(-n\Bnew GR^{-\alpha}\right)\left(GR^{-\alpha}\right)^{t_i}$ is defined for analytical convenience.
\begin{figure*}[!t]
\normalsize
\begin{align}\label{ana_outage_form}
\epsilon =1-\sum_{n\in \mathcal{N}}\exp\left(-n\A\right)\sum_{k\in \mathcal{K}}a_{nk}\sum_{j=0}^{k}\binom{k}{j}\left(\A\right)^{k-j}\left(\Bnew\right)^j\sum_{t_1+t_2...+t_M=j}\binom{j}{t_1,t_2,...,t_M}\prod_{i=1}^{M}\mathrm{E}_{G,R}\left\{\E\right\}.
\end{align}
\hrulefill
\vspace*{4pt}
\end{figure*}
\fi

\ifCLASSOPTIONonecolumn
Using assumption A4, the outage probability in~\eqref{ana_outage_form} can be expressed for the case of Nakagami-$m$ fading with integer $m_0$ as
\begin{align}\label{ana_gamma_outage}
\epsilon =&1-\exp\left(-m_0\A\right)\sum_{k=0}^{m_0-1}\frac{m_0^k}{k!}\sum_{j=0}^{k}\binom{k}{j}\left(\A\right)^{k-j}\left(\Bnew\right)^j \sum_{t_1+t_2...+t_M=j}\binom{j}{t_1,t_2,...,t_M}\prod_{i=1}^{M}\mathrm{E}_{G,R}\left\{\E\right\},
\end{align}

\noindent where the expectation in~\eqref{ana_gamma_outage} can be expressed using~\eqref{g_i_distri} as
\begin{align}\label{ana_new_expt}
\mathrm{E}_{G,R}\left\{\E\right\}=\int_0^{r_{\textrm{max}}}\frac{m^m(r^{-\alpha})^{t_i}\Gamma(m+t_i)}{\Gamma(m)\left(m+\Bnew m_0r^{-\alpha}\right)^{m+t_i}}\fr dr,
\end{align}

\noindent where $r_{\textrm{max}}$ is defined below~\eqref{LT_gamma_z}.

\else
Using assumption A4, the outage probability in~\eqref{ana_outage_form} can be expressed for the case of Nakagami-$m$ fading with integer $m_0$ as~\eqref{ana_gamma_outage}, shown at the top of the next page, where the expectation in~\eqref{ana_gamma_outage} can be expressed using~\eqref{g_i_distri} as
\begin{figure*}[!t]
\normalsize
\begin{align}
\label{ana_gamma_outage}
\epsilon =&1-\exp\left(-m_0\A\right)\sum_{k=0}^{m_0-1}\frac{m_0^k}{k!}\sum_{j=0}^{k}\binom{k}{j}\left(\A\right)^{k-j}\left(\Bnew\right)^j \sum_{t_1+t_2...+t_M=j}\binom{j}{t_1,t_2,...,t_M}\prod_{i=1}^{M}\mathrm{E}_{G,R}\left\{\E\right\}.
\end{align}
\hrulefill
\vspace*{4pt}
\end{figure*}
\begin{align}\label{ana_new_expt}
\mathrm{E}_{G,R}\left\{\E\right\}=\int_0^{r_{\textrm{max}}}\frac{m^m(r^{-\alpha})^{t_i}\Gamma(m+t_i)}{\Gamma(m)\left(m+\Bnew m_0r^{-\alpha}\right)^{m+t_i}}\fr dr,
\end{align}

\noindent where $r_{\textrm{max}}$ is defined below~\eqref{LT_gamma_z}.
\fi
\begin{remark}\label{remark2}
For the RLPG-based framework, $m_0$ is constrained to take integer values only, while $m$ can take any value. The restriction on $m_0$ is because $m_0-1$, as an upper limit for the summation in~\eqref{ana_gamma_outage}, can only be integer. This is in contrast with~\eqref{final_gama_outage} where both $m_0$ and $m$ can take any (integer or non-integer) value for the MGF-based framework.
\end{remark}

Summarizing,~\eqref{final_outage} and~\eqref{ana_outage_form1} take the form of~\eqref{final_gama_outage} and~\eqref{ana_gamma_outage}, respectively, for $M$ interfering nodes i.i.d. at random inside an arbitrarily-shaped finite wireless network with i.i.d. Nakagami-$m$ fading channels.

\ifCLASSOPTIONonecolumn
\vspace{-0.35 cm}
\fi

\subsection{Need for the Two Frameworks}
The proposed two frameworks complement each other. On one hand, as highlighted in Remark~\ref{remark2}, the MGF-based framework can be used in scenarios with non-integer $m_0$ while the RLPG-based framework can only be used in scenarios with integer $m_0$. On the other hand, the RLPG-based framework is more capable of yielding closed-form analytical expressions than the MGF-based framework. For the MGF-based framework, the outage probability in~\eqref{final_gama_outage} involves a double integration (an integration with an expectation term in the integrand). In general, it is not possible to obtain a closed-form for the integration part in~\eqref{final_gama_outage} because the expectation term is raised to a power factor of $M$ ($M\geq2$). However, in certain cases, the expectation term can be expressed in closed-form. For the RLPG-based framework, the outage probability in~\eqref{ana_gamma_outage} involves a single integration which admits closed-form results in a much larger number of cases.

Note that both~\eqref{final_gama_outage} and~\eqref{ana_gamma_outage} require the knowledge of the distance distribution $f_R(r)$, i.e. the PDF of the distance of a random node from the reference receiver $Y_0$, for their evaluation. The distance distribution $\fr$ is dependent on the underlying random model for the node locations. For the uniform BPP (assumption A5), which is considered in this work, the distance distribution $\fr$ is derived in~\cite{Srinivasa-2010} for the special case when the reference receiver is located at the center of a convex regular polygon. Recently, the result in~\cite{Srinivasa-2010} was generalized in~\cite{Zubair-2013} for the case when the reference receiver is located anywhere inside a convex regular polygon. It was shown in~\cite{Zubair-2013} that for an arbitrary location of the reference receiver inside convex regular polygon, the distance distribution $\fr$ can be a complicated piece-wise function because of the boundary effects. We note that the approach in~\cite{Zubair-2013} is also applicable for arbitrarily-shaped convex polygons (see Appendix~\ref{appendix:fr} for details). Once $\fr$ is given, both~\eqref{final_gama_outage} and~\eqref{ana_gamma_outage} and can be accurately evaluated.

In the next two sections, we show how~\eqref{final_gama_outage} and~\eqref{ana_gamma_outage} can be evaluated in disk and polygon regions, which are commonly used in the literature for the modeling of wireless networks. We also illustrate how the proposed frameworks can be applied in the case of arbitrarily-shaped finite wireless networks with arbitrary location of the reference receiver.


\section{Outage Probability in a Disk Region}\label{disk}
Consider the scenario that the region $\area$ is a disk of radius $\radius$, as shown in Fig.~\ref{diskfig:one}. The reference receiver is assumed to be located at a distance $d$ from the center of the disk. Then, the distance distribution $\fr$ can be exactly expressed as~\cite{Zubair-2013}
\ifCLASSOPTIONonecolumn
\begin{align}\label{f_r_disk}
\fr= \frac{1}{|\area|}\left\{ \begin{array}{ll}
       2\pi r, &{0\leq r \leq \radius-d ;}\\
       2r\arccos\left(\frac{r^2+d^2-\radius^2}{2dr}\right), \!\!\! &{\radius-d\leq r \leq\radius+d;}
                    \end{array} \right.
\end{align}
\else
\begin{align}\label{f_r_disk}
{\small
\fr= \frac{1}{|\area|}\left\{ \begin{array}{ll}
       2\pi r, &{0\leq r \leq \radius-d ;}\\
       2r\arccos\left(\frac{r^2+d^2-\radius^2}{2dr}\right), \!\!\! &{\radius-d\leq r \leq\radius+d;}
                    \end{array} \right.}
\end{align}
\fi

\noindent Note that substituting $d=0$ in~\eqref{f_r_disk} gives the distance distribution for the special case that $Y_0$ is located at the center of the disk. Similarly, substituting $d=\radius$ in~\eqref{f_r_disk} gives the distance distribution for the special case that $Y_0$ is located anywhere on the circumference of the disk.
%
\ifCLASSOPTIONonecolumn
\begin{figure}[t]
 \centering
    \includegraphics[width=0.35\textwidth]{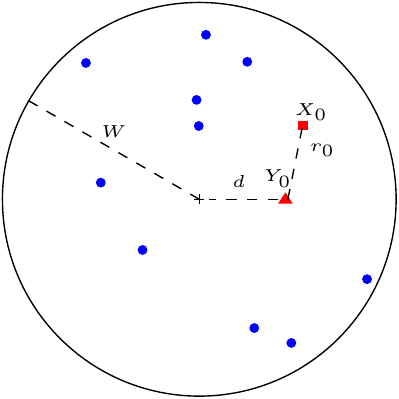}
\caption{Illustration of a finite wireless network with arbitrary location of reference receiver $Y_0$ in a disk region of radius $\radius$ ($+=$ center of disk, ${\color{blue}\bullet}=$ interfering node, ${\color{red}\blacktriangle}=$ reference receiver, ${\color{red}\blacksquare}=$ reference transmitter).}  \label{diskfig:one}
\end{figure}
\else
\begin{figure}[t]
 \centering
    \includegraphics[width=0.3\textwidth]{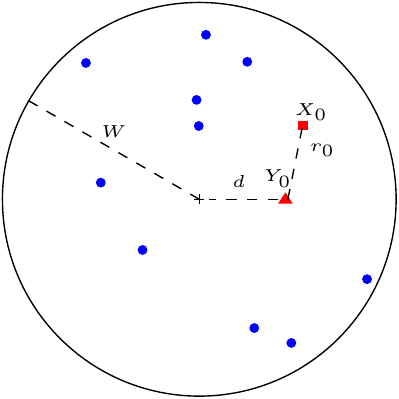}
\caption{Illustration of a finite wireless network with arbitrary location of reference receiver $Y_0$ in a disk region of radius $\radius$ ($+=$ center of disk, ${\color{blue}\bullet}=$ interfering node, ${\color{red}\blacktriangle}=$ reference receiver, ${\color{red}\blacksquare}=$ reference transmitter).}  \label{diskfig:one}
\end{figure}
\fi

\subsection{MGF-based Framework}

Substituting~\eqref{f_r_disk} in~\eqref{LT_gamma_z}, we find that the expectation has a closed-form only for the first part of the range $(0\leq r\leq\radius-d)$ in~\eqref{f_r_disk}. For the second-part of the range $(\radius-d\leq r\leq\radius+d)$, the integration does not have a closed-form due to the $\arccos(\cdot)$ function.
%
\ifCLASSOPTIONonecolumn
The result is shown in~\eqref{mgf_disk_expectation}.
\begin{align}\label{mgf_disk_expectation}
\mathrm{E}_{G,R}\left\{\exp\left(\frac{-sGR^{-\alpha}}{r_0^{-\alpha}G_0}\right)\right\}=\varphi\left(2\pi, \radius-d\right)+\frac{2m^m}{|\area| }\int_{\radius-d}^{\radius+d}\left(m+\frac{r^{-\alpha}r_0^{\alpha}s}{g_0}\right)^{-m}r \arccos\left(\frac{r^2+d^2-\radius^2}{2dr}\right) dr,
\end{align}
\else
The result is shown in~\eqref{mgf_disk_expectation} at the top of this page, where $\varphi(\cdot,\cdot)$ is defined in~\eqref{eq:varphi} also at the top of this page.
\begin{figure*}[!t]
\normalsize
\begin{align}\label{mgf_disk_expectation}
\mathrm{E}_{G,R}\left\{\exp\left(\frac{-sGR^{-\alpha}}{r_0^{-\alpha}G_0}\right)\right\}=\varphi\left(2\pi, \radius-d\right)+\frac{2m^m}{|\area| }\int_{\radius-d}^{\radius+d}\left(m+\frac{r^{-\alpha}r_0^{\alpha}s}{g_0}\right)^{-m}r \arccos\left(\frac{r^2+d^2-\radius^2}{2dr}\right) dr.
\end{align}
\hrulefill
\vspace*{4pt}
\end{figure*}
\fi
%
\ifCLASSOPTIONonecolumn
where
\begin{align}\label{eq:varphi}
\varphi\left(\theta, \upsilon\right)=&\frac{\theta m^m g_0^m\upsilon^{2+\alpha m}}{|\area|(2+\alpha m)\left(r_0^\alpha s\right)^m} \,_2F_1\left[m,\frac{2}{\alpha}+m,1+\frac{2}{\alpha}+m,-\frac{g_0 m \upsilon^{\alpha}}{r_0^{\alpha}s} \right].
\end{align}
\else
\begin{figure*}[!t]
\normalsize
\begin{align}\label{eq:varphi}
\varphi\left(\theta, \upsilon\right)=&\frac{\theta m^m g_0^m\upsilon^{2+\alpha m}}{|\area|(2+\alpha m)\left(r_0^\alpha s\right)^m} \,_2F_1\left[m,\frac{2}{\alpha}+m,1+\frac{2}{\alpha}+m,-\frac{g_0 m \upsilon^{\alpha}}{r_0^{\alpha}s} \right].
\end{align}
\hrulefill
\vspace*{4pt}
\end{figure*}
\fi
%
Substituting~\eqref{mgf_disk_expectation} in~\eqref{final_gama_outage}, the outage probability can be numerically evaluated.

\subsection{RLPG-based Framework}

\ifCLASSOPTIONonecolumn
Substituting~\eqref{f_r_disk} in~\eqref{ana_new_expt} and after some manipulations, we get
\begin{align}\label{e_disk}
\mathrm{E}_{G,R}\left\{\E \right\}=&\psi\left(2\pi, \radius-d, t_i\right)+\frac{2m^m\Gamma(m+t_i)}{|\area| \Gamma(m)}\int_{\radius-d}^{\radius+d}\frac{(r^{-\alpha})^{t_i}}{\left(m+\Bnew m_0r^{-\alpha}\right)^{m+t_i}}r\arccos\left(\frac{r^2+d^2-\radius^2}{2dr}\right)dr,
\end{align}
\else
Substituting~\eqref{f_r_disk} in~\eqref{ana_new_expt} and after some manipulations, we get~\eqref{e_disk} shown at the top of the next page, where $\psi(\cdot,\cdot,\cdot)$ is defined in~\eqref{eq:psi} also at the top of the next page.
\begin{figure*}[!t]
\normalsize
\begin{align}\label{e_disk}
\mathrm{E}_{G,R}\left\{\E \right\}=&\psi\left(2\pi, \radius-d, t_i\right)+\frac{2m^m\Gamma(m+t_i)}{|\area| \Gamma(m)}\int_{\radius-d}^{\radius+d}\frac{(r^{-\alpha})^{t_i}}{\left(m+\Bnew m_0r^{-\alpha}\right)^{m+t_i}}r\arccos\left(\frac{r^2+d^2-\radius^2}{2dr}\right)dr.
\end{align}
\hrulefill
\vspace*{4pt}
\end{figure*}
\fi
\ifCLASSOPTIONonecolumn
where
\begin{align}\label{eq:psi}
\psi\left(\theta, \upsilon, \tau\right)=\frac{\theta m^m\left(\Bnew m_0\right)^{-m-\tau}\upsilon^{2+\alpha m}\Gamma(m+\tau)}{|\area|(2+\alpha m)\Gamma(m)}   \,_2F_1\left[\frac{2}{\alpha}+m,m+\tau,1+\frac{2}{\alpha}+m,-\frac{m\upsilon^{\alpha}}{\Bnew m_0} \right].
\end{align}
\else
\begin{figure*}[!t]
\normalsize
\begin{align}\label{eq:psi}
\psi\left(\theta, \upsilon, \tau\right)=\frac{\theta m^m\left(\Bnew m_0\right)^{-m-\tau}\upsilon^{2+\alpha m}\Gamma(m+\tau)}{|\area|(2+\alpha m)\Gamma(m)}   \,_2F_1\left[\frac{2}{\alpha}+m,m+\tau,1+\frac{2}{\alpha}+m,-\frac{m\upsilon^{\alpha}}{\Bnew m_0} \right].
\end{align}
\hrulefill
\vspace*{4pt}
\end{figure*}
\fi
%
In general, the integration in~\eqref{e_disk} also does not have a closed-form due to the $\arccos(\cdot)$ function. It is possible to use the Gauss-Chebyshev integration technique~\cite{Hildebrand1974} in order to obtain an approximate closed-form expression. However, in our investigations, we found that a summation over a large number of terms $(>1000)$ was required in our case for accurate evaluation. Hence, we do not pursue approximations and instead directly substitute~\eqref{e_disk} in~\eqref{ana_gamma_outage} to obtain the outage probability.

\noindent \textit{\underline{Special case:}} For the case of the reference receiver $Y_0$ located at the center of the disk, $d=0$. Substituting this value in~\eqref{e_disk} and then substituting the result in~\eqref{ana_gamma_outage}, the final expression for the outage probability simplifies to
\ifCLASSOPTIONonecolumn
\begin{align}\label{ana_disk_center}
\epsilon_{\textrm{center}} =1-\exp\left(-m_0\A\right)\sum_{k=0}^{m_0-1}\frac{m_0^k}{k!}\sum_{j=0}^{k}\binom{k}{j}\left(\A\right)^{k-j}\left(\Bnew\right)^j \sum_{t_1+t_2...+t_M=j}\binom{j}{t_1,t_2,...,t_M}\prod_{i=1}^{M}\psi\left(2\pi, \radius, t_i\right),
\end{align}
\else
\begin{align}\label{ana_disk_center}
&\epsilon_{\textrm{center}} =1-\exp\left(\frac{-m_0\beta}{\rho_0}\right)\sum_{k=0}^{m_0-1}\frac{m_0^k}{k!}\sum_{j=0}^{k}\binom{k}{j}\left(\A\right)^{k-j} \times\nonumber\\
&\left(\Bnew\right)^j  \!\!\!\sum_{t_1+t_2...+t_M=j} \binom{j}{t_1,t_2,...,t_M}\prod_{i=1}^{M}\psi\left(2\pi, \radius, t_i\right),
\end{align}
\fi
\noindent where $\psi(\cdot,\cdot,\cdot)$ is defined in~\eqref{eq:psi}.
\begin{remark}\label{remark3}
The outage probability for a finite number of nodes i.u.d. in a disk region has been widely considered in the recent literature. Our proposed frameworks reproduce the available outage results in the literature as special cases. For the MGF-based framework, with $m_0=1$, the result from~\eqref{mgf_disk_expectation} is equivalent to the result in~\cite[eq.(24)]{Srinivasa-2007}. For the RLPG-based framework, with the reference receiver located at the center of the network,~\eqref{ana_disk_center} is identical to the result in~\cite[eq.(44)]{Torrieri-2012}.
\end{remark}

\section{Outage Probability in Regular Polygons and Arbitrarily-Shaped Convex Polygons}\label{other}
In this section, we illustrate the exact computation of the outage probability in both regular and arbitrarily-shaped convex polygon regions. We will consider the following two cases (i) reference receiver $Y_0$ located at the center of a regular $L$-sided polygon and (ii) reference receiver $Y_0$ located at an arbitrary location in an arbitrarily-shaped region.

\subsection{Center of Polygon}
Consider the finite region $\area$ to be a regular $L$-sided convex polygon which is inscribed in a circle of radius $\radius$. Then, the area and the interior angle between two adjacent sides are given by
 \begin{subequations}
 \begin{align}
 |\area|&=\frac{1}{2}L\radius^2 \sin\left(\frac{2\pi}{L}\right),\\
  \theta &=\frac{\pi(L-2)}{L}.
 \end{align}
 \end{subequations}

In general, polygon regions are non-isotropic. Hence, there is no single expression for the distance distribution $\fr$ for an arbitrary location of the reference receiver inside a convex regular polygon. For the special case that the reference receiver $Y_0$ is located at the center of an $L$-sided convex regular polygon, the distance distribution $\fr$ can be expressed as~\cite{Srinivasa-2010,Zubair-2013}
\begin{align}\label{polygon_center_pdf}
\fr=\frac{1}{|\area|}  \left\{ \begin{array}{ll}
        2\pi r, &{0\leq r\leq \radius\sin{\left(\frac{\theta}{2}\right)};}\\
      2\pi r -2Lr \Delta,                 &{\radius\sin{\left(\frac{\theta}{2}\right)}\leq r\leq\radius;}
                    \end{array} \right.
\end{align}

\noindent where $\Delta=\arccos\left(\frac{\radius\sin{\left(\frac{\theta}{2}\right)}}{r} \right)$. Using~\eqref{polygon_center_pdf}, we illustrate the computation of the outage probability using the two frameworks.

\noindent \underline{\textit{MGF-based Framework:}}
%
\ifCLASSOPTIONonecolumn
Substituting~\eqref{polygon_center_pdf} in~\eqref{LT_gamma_z} and after some manipulations, we get
\begin{align}\label{mgf_center_poly}
\mathrm{E}_{G,R}\left\{\exp\left(\frac{-sGR^{-\alpha}}{r_0^{-\alpha}G_0}\right)\right\}=&\varphi\left(2\pi,\radius\right)-\frac{2Lm^m}{|\area|}\int_{\radius\sin{\left(\frac{\theta}{2}\right)}}^{\radius}\left(m+\frac{r^{-\alpha}r_0^{\alpha}s}{g_0}\right)^{-m}r \arccos\left(\frac{\radius\sin{\left(\frac{\theta}{2}\right)}}{r}\right) dr,
\end{align}
\else
\noindent Substituting~\eqref{polygon_center_pdf} in~\eqref{LT_gamma_z} and after some manipulations, we get~\eqref{mgf_center_poly} shown at the top of the next page,
\begin{figure*}[!t]
\normalsize
\begin{align}\label{mgf_center_poly}
\mathrm{E}_{G,R}\left\{\exp\left(\frac{-sGR^{-\alpha}}{r_0^{-\alpha}G_0}\right)\right\}=\varphi\left(2\pi,\radius\right)-\frac{2Lm^m}{|\area|}\int_{\radius\sin{\left(\frac{\theta}{2}\right)}}^{\radius}\left(m+\frac{r^{-\alpha}r_0^{\alpha}s}{g_0}\right)^{-m}r \arccos\left(\frac{\radius\sin{\left(\frac{\theta}{2}\right)}}{r}\right) dr.
\end{align}
\hrulefill
\vspace*{4pt}
\end{figure*}
\fi
\noindent where $\varphi(\cdot,\cdot)$ is defined in~\eqref{eq:varphi}. Substituting~\eqref{mgf_center_poly} in~\eqref{final_gama_outage}, the outage probability can be evaluated.

\noindent \underline{\textit{RLPG-based Framework:}}
%
\ifCLASSOPTIONonecolumn
Substituting~\eqref{polygon_center_pdf} in~\eqref{ana_new_expt}, and after some manipulations, we get
\begin{align}\label{ana_center_e}
\mathrm{E}_{G,R}\left\{\E\right\}=\psi\left(2\pi,\radius, t_i\right)-\frac{2Lm^m\Gamma(m+t_i)}{|\area|\Gamma(m)}\int_{\radius\sin{\left(\frac{\theta}{2}\right)}}^{\radius}\frac{(r^{-\alpha})^{t_i}}{\left(m+\Bnew m_0r^{-\alpha}\right)^{m+t_i}}r\arccos\left(\frac{\radius\sin{\left(\frac{\theta}{2}\right)}}{r} \right) dr,
\end{align}
\else
Substituting~\eqref{polygon_center_pdf} in~\eqref{ana_new_expt}, and after some manipulations, we get~\eqref{ana_center_e} shown at the top of the next page,
\begin{figure*}[!t]
\normalsize
\begin{align}\label{ana_center_e}
\mathrm{E}_{G,R}\left\{\E\right\}=\psi\left(2\pi,\radius, t_i\right)-\frac{2Lm^m\Gamma(m+t_i)}{|\area|\Gamma(m)}\int_{\radius\sin{\left(\frac{\theta}{2}\right)}}^{\radius}\frac{(r^{-\alpha})^{t_i}}{\left(m+\Bnew m_0r^{-\alpha}\right)^{m+t_i}}r\arccos\left(\frac{\radius\sin{\left(\frac{\theta}{2}\right)}}{r} \right) dr.
\end{align}
\hrulefill
\vspace*{4pt}
\end{figure*}
\fi
\noindent where $\psi(\cdot,\cdot,\cdot)$ is defined in~\eqref{eq:psi}. Substituting~\eqref{ana_center_e} in~\eqref{ana_gamma_outage}, the outage probability can be evaluated.

\subsection{Arbitrarily-Shaped Convex Polygon Region}
We consider an arbitrarily-shaped convex polygon region as shown in Fig.~\ref{fig:two}, with side lengths $S_1=S_2=\sqrt{3}\radius$, $S_3=\sqrt{7-3\sqrt{3}-\sqrt{6}}\radius$ and $S_4=\radius$. Thus, the interior angles formed at vertices $V_1$, $V_2$, $V_3$ and $V_4$ are $\pi/2$, $\pi/4$, $0.6173\pi$ and $0.6327\pi$. Suppose that $Y_0$ is located at the vertex $V_2$\footnote{The $Y_0$ location at vertex $V_2$ is chosen here for the sake of simplicity. Later in Section VII, we also show results for an arbitrary location of $Y_0$ inside the arbitrarily-shaped convex polygon region considered in Fig.~2.}. Then, following the derivation in Appendix~\ref{appendix:fr}, $\fr$ can be expressed as
%
\ifCLASSOPTIONonecolumn
\begin{align}\label{arb_shape_pdf}
\fr= \frac{1}{|\area|}\left\{\begin{array}{ll}
        \frac{\pi}{4}r, &{0\leq r\leq\sqrt{3}\radius;}\\
         0.3673\pi r-r\arccos\left(\frac{\sqrt{3}\radius}{r}\right)-r\arccos\left(\frac{1.6159\radius}{r}\right), &{\sqrt{3}\radius\leq r\leq2\radius;}
                    \end{array} \right.
\end{align}
\else
\begin{align}\label{arb_shape_pdf}
\fr= \frac{1}{|\area|}\left\{\begin{array}{ll}
        \frac{\pi}{4}r, & \! \! \! \! \! \!{0\leq r\leq\sqrt{3}\radius;}\\
         0.3673\pi r-r\arccos\left(\frac{\sqrt{3}\radius}{r}\right)\\
         -r\arccos\left(\frac{1.6159\radius}{r}\right), & \!\!\!\!\!\!\!\!\!\!\!\!\!{\sqrt{3}\radius\leq r\leq2\radius;}
                    \end{array} \right.
\end{align}
\fi

\noindent Using~\eqref{arb_shape_pdf}, we illustrate the computation of the outage probability using the two frameworks.

\noindent \underline{\textit{MGF-based Framework:}}
\ifCLASSOPTIONonecolumn
Substituting~\eqref{arb_shape_pdf} in~\eqref{LT_gamma_z}, the expectation can be expressed as
\begin{align}\label{arb_mgf}
\mathrm{E}_{G,R}\left\{\exp\left(\frac{-sGR^{-\alpha}}{r_0^{-\alpha}G_0}\right)\right\}=&\varphi\left(0.3673\pi,2\radius\right)-\varphi\left(0.1173\pi,\sqrt{3}\radius\right)\nonumber\\
&-\frac{m^m}{|\area|}\int_{\sqrt{3}\radius}^{2\radius} \left(m+\frac{r^{-\alpha}r_0^{\alpha}s}{g_0}\right)^{-m}r\left(\arccos\left(\frac{\sqrt{3}\radius}{r}\right)+\arccos\left(\frac{1.6159\radius}{r}\right)\right)dr,
\end{align}
\else
Substituting~\eqref{arb_shape_pdf} to~\eqref{LT_gamma_z}, the expectation can be expressed as~\eqref{arb_mgf}, shown at the top of the next page,
\begin{figure*}[!t]
\normalsize
\begin{align}\label{arb_mgf}
\mathrm{E}_{G,R}\left\{\exp\left(\frac{-sGR^{-\alpha}}{r_0^{-\alpha}G_0}\right)\right\}=&\varphi\left(0.3673\pi,2\radius\right)-\varphi\left(0.1173\pi,\sqrt{3}\radius\right)\nonumber\\
&-\frac{m^m}{|\area|}\int_{\sqrt{3}\radius}^{2\radius} \left(m+\frac{r^{-\alpha}r_0^{\alpha}s}{g_0}\right)^{-m}r\left(\arccos\left(\frac{\sqrt{3}\radius}{r}\right)+\arccos\left(\frac{1.6159\radius}{r}\right)\right)dr.
\end{align}
\hrulefill
\vspace*{4pt}
\end{figure*}
\fi
%
\noindent where $\varphi(\cdot,\cdot)$ is defined in~\eqref{eq:varphi}. Finally, the outage probability for the case of the reference receiver located at vertex $V_2$ can be evaluated by substituting~\eqref{arb_mgf} in~\eqref{final_gama_outage}.

\noindent \underline{\textit{RLPG-based Framework:}}
%
\ifCLASSOPTIONonecolumn
Substituting~\eqref{arb_shape_pdf} in~\eqref{ana_new_expt}, the expectation can be expressed as
\begin{align}\label{final_ana_e}
\mathrm{E}_{G,R}\left\{\E\right\}=&\psi\left(0.3673\pi,2\radius, t_i\right)-\psi\left(0.1173\pi,\sqrt{3}\radius, t_i\right)\nonumber\\
&-\frac{m^m\Gamma(m+t_i)}{|\area|\Gamma(m)}\int_{\sqrt{3}\radius}^{2\radius}\frac{(r^{-\alpha})^{t_i}}{\left(m+\Bnew m_0r^{-\alpha}\right)^{m+t_i}}r\left(\arccos\left(\frac{\sqrt{3}\radius}{r}\right)+\arccos\left(\frac{1.6159\radius}{r}\right)\right) dr,
\end{align}
\else
Substituting~\eqref{arb_shape_pdf} in~\eqref{ana_new_expt}, the expectation can be expressed as~\eqref{final_ana_e} shown at the top of the next page,
\begin{figure*}[!t]
\normalsize
\begin{align}\label{final_ana_e}
\mathrm{E}_{G,R}\left\{\E\right\}=&\psi\left(0.3673\pi,2\radius, t_i\right)-\psi\left(0.1173\pi,\sqrt{3}\radius, t_i\right)\nonumber\\
&-\frac{m^m\Gamma(m+t_i)}{|\area|\Gamma(m)}\int_{\sqrt{3}\radius}^{2\radius}\frac{(r^{-\alpha})^{t_i}}{\left(m+\Bnew m_0r^{-\alpha}\right)^{m+t_i}}r\left(\arccos\left(\frac{\sqrt{3}\radius}{r}\right)+\arccos\left(\frac{1.6159\radius}{r}\right)\right) dr.
\end{align}
\hrulefill
\vspace*{4pt}
\end{figure*}
\fi
%
\noindent where $\psi(\cdot,\cdot,\cdot)$ is defined in~\eqref{eq:psi}. Finally, the outage probability for the case of the reference receiver located at vertex $V_2$ can be evaluated by substituting~\eqref{final_ana_e} in~\eqref{ana_gamma_outage}.

Summarizing, the procedure for deriving the outage probability for i.u.d. nodes in an arbitrarily-shaped convex polygon region with i.i.d. Nakagami-$m$ fading channels is summarized in Algorithm~\ref{algo:2}.
\begin{algorithm}[h]
\caption{Proposed Algorithm} \label{algo:2}
\begin{algorithmic}
\STATE Step 1: Choose the location of the reference receiver inside the arbitrarily-shaped convex polygon region.
\STATE Step 2: Determine $\fr$ based on the approach summarized in Appendix~\ref{appendix:fr}.
\STATE Step 3: Depending on the value of $m_0$, select the appropriate framework to calculate the outage probability.
\IF{$m_0$ is non-integer} \STATE
Use the MGF-based framework. Substitute $\fr$ in~\eqref{LT_gamma_z} and then~\eqref{LT_gamma_z} in~\eqref{final_gama_outage} to compute the outage probability.   \ELSE \STATE
Use the RLPG-based framework. Substitute $\fr$ in~\eqref{ana_new_expt} and then~\eqref{ana_new_expt} in~\eqref{ana_gamma_outage} to compute the outage probability.
\ENDIF
\end{algorithmic}
\end{algorithm}

\ifCLASSOPTIONonecolumn
\begin{figure}[t]
 \centering
    \includegraphics[width=0.5\textwidth]{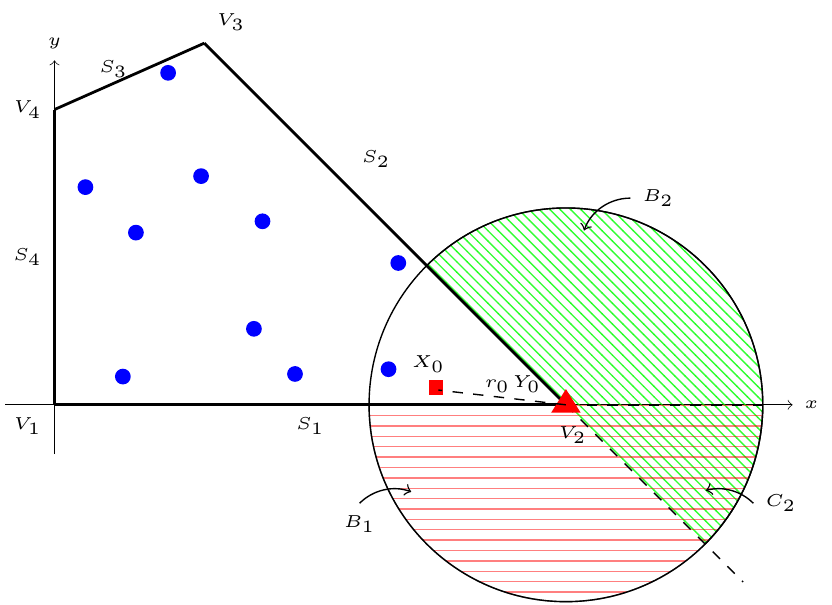}
    \caption{Illustration of an arbitrary location of a reference receiver in an arbitrarily-shaped finite wireless network, with side lengths $S_1=\sqrt{3}\radius$, $S_2=\sqrt{3}\radius$,  $S_3=\sqrt{7-3\sqrt{3}-\sqrt{6}}\radius$ and $S_4=\radius$ and vertices $V_1$, $V_2$, $V_3$ and $V_4$. The areas $B_1$ (shaded in horizontal lines), lines) and $C_2$ (intersection of horizontal and diagonal lines) are defined in Appendix~\ref{appendix:fr} (${\color{blue}\bullet}=$ interfering node, ${\color{red}\blacktriangle}=$ reference receiver, ${\color{red}\blacksquare}=$ reference transmitter).}  \label{fig:two}
\end{figure}
\else
\begin{figure}[t]
 \centering
    \includegraphics[width=0.4\textwidth]{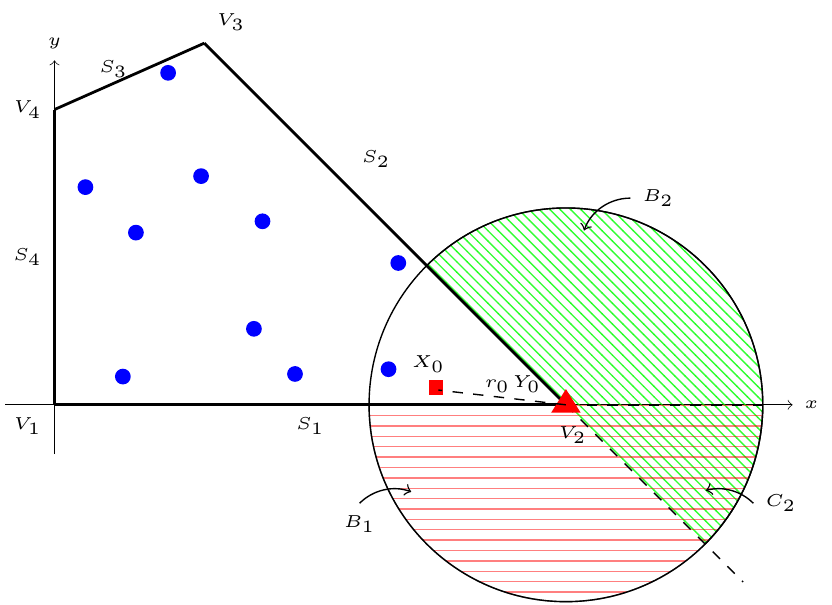}
    \caption{Illustration of an arbitrary location of a reference receiver in an arbitrarily-shaped finite wireless network, with side lengths $S_1=\sqrt{3}\radius$, $S_2=\sqrt{3}\radius$,  $S_3=\sqrt{7-3\sqrt{3}-\sqrt{6}}\radius$ and $S_4=\radius$ and vertices $V_1$, $V_2$, $V_3$ and $V_4$. The areas $B_1$ (shaded in horizontal lines), lines) and $C_2$ (intersection of horizontal and diagonal lines) are defined in Appendix~\ref{appendix:fr} (${\color{blue}\bullet}=$ interfering node, ${\color{red}\blacktriangle}=$ reference receiver, ${\color{red}\blacksquare}=$ reference transmitter).}  \label{fig:two}
\end{figure}
\fi

\section{Numerical and Simulation Results}\label{result}
In this section, we first address the computational aspects of the two frameworks. We then study the outage probability performance of arbitrarily-shaped finite wireless networks and discuss the boundary effects in finite wireless networks in detail.

\subsection{Computational Aspects of the Frameworks}\label{compute}

In general, both frameworks require numerical evaluation of integration, for which any standard mathematical package such as Matlab or Mathematica can be used. It must be noted that the numerical evaluation of single and double integrations is standard and widely practiced in the wireless communications literature~\cite{Beaulieu2012}.

For the MGF-based framework, the outage expressions in~\eqref{final_outage},~\eqref{final_outage1} and~\eqref{final_gama_outage} are a summation over a finite number of terms. The three parameters $A$, $B$ and $C$ (defined below~\eqref{LT_main}) control the estimation error. Following the well established guidelines in~\cite{Abate-1995},~\cite{OCinneide-1997}, in order to achieve an estimation accuracy of $10^{-\zeta}$ (i.e., having the $\zeta-1$th decimal correct), $A$, $B$ and $C$ have to at least equal $\zeta \ln 10$, $1.243 \zeta -1$, and $1.467\zeta$, respectively. For example, for the disk region, we set $A = 8 \ln 10$, $B = 11$, $C = 14$. This achieves stable numerical inversion with an estimation error of $10^{-8}$.

For the RLPG-based framework, if both $m_0$ and $M$ are large, then the computation of all possible integer results for $t_i$ in $t_1+t_2+...+t_M=j$ $(j=0,...,m_0-1)$ in~\eqref{ana_outage_form1},~\eqref{ana_outage_form} and~\eqref{ana_gamma_outage} can be time-consuming. This is due to the fact that we need to use $M$ \textsf{for loops} to find the complete results. However, when either $m_0$ or $M$ is a small number we can pre-compute these results, as suggested in~\cite{Torrieri-2012}, and store them as a matrix for use in computations.

%

\subsection{Validation of the Proposed Two Frameworks}
%
\ifCLASSOPTIONonecolumn
\begin{figure}[t]
\centering
        \includegraphics[width=0.68  \textwidth]{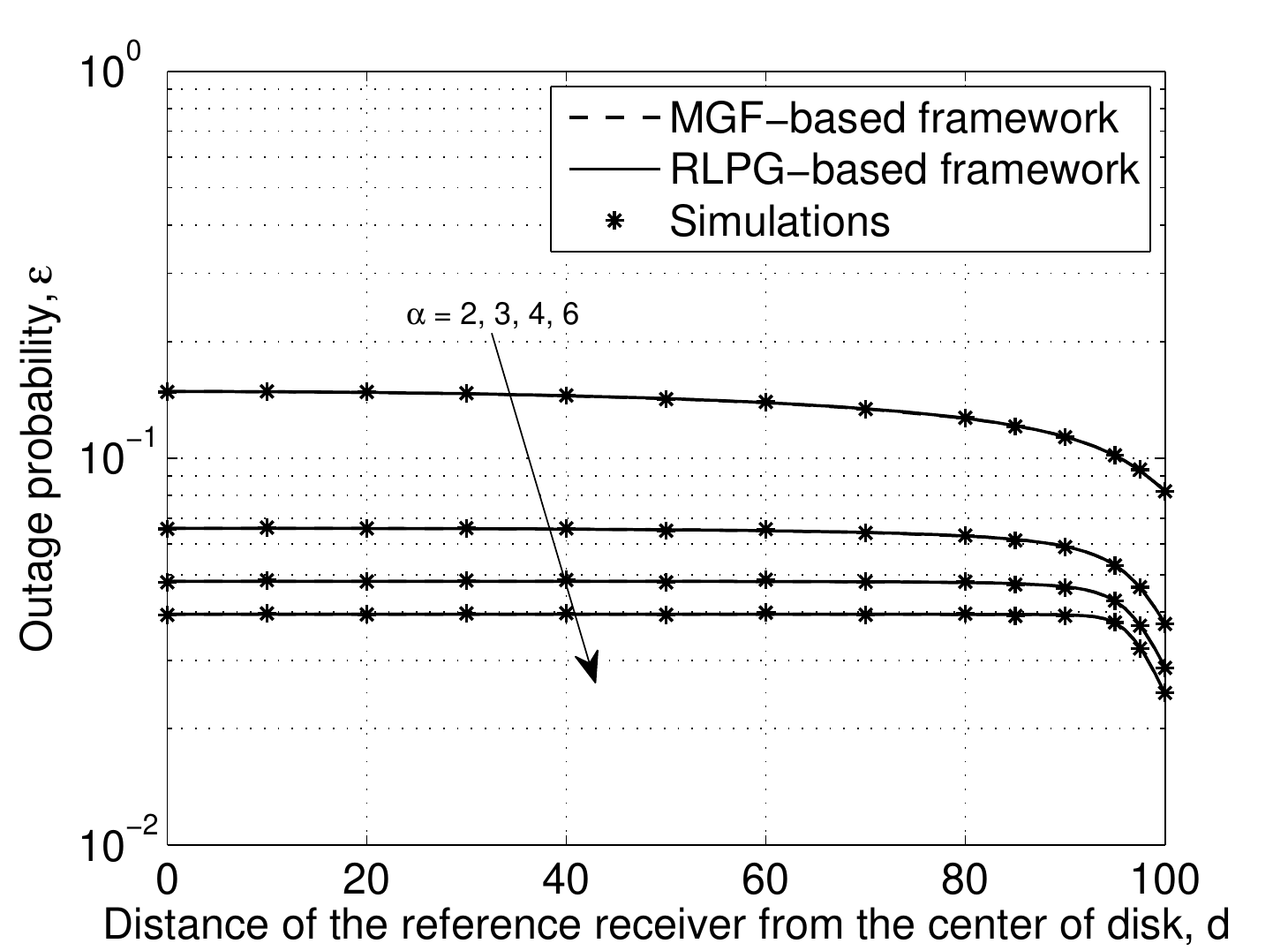}
        \caption{Outage probability, $\epsilon$, versus the distance of the reference receiver from the center of disk, $d$, for $M=10$ interferers i.u.d. in a disk of radius $\radius=100$, with reference link distance $r_0=5$, path-loss exponents $\alpha=2, 3, 4, 6$, i.i.d. Rayleigh fading channels ($m_0=m=1$), SINR threshold $\beta=0$ dB and SNR $\rho_0=20$ dB.}
        \label{disk_fig1}
\end{figure}
\else
\begin{figure}[t]
\centering
        \includegraphics[width=0.5  \textwidth]{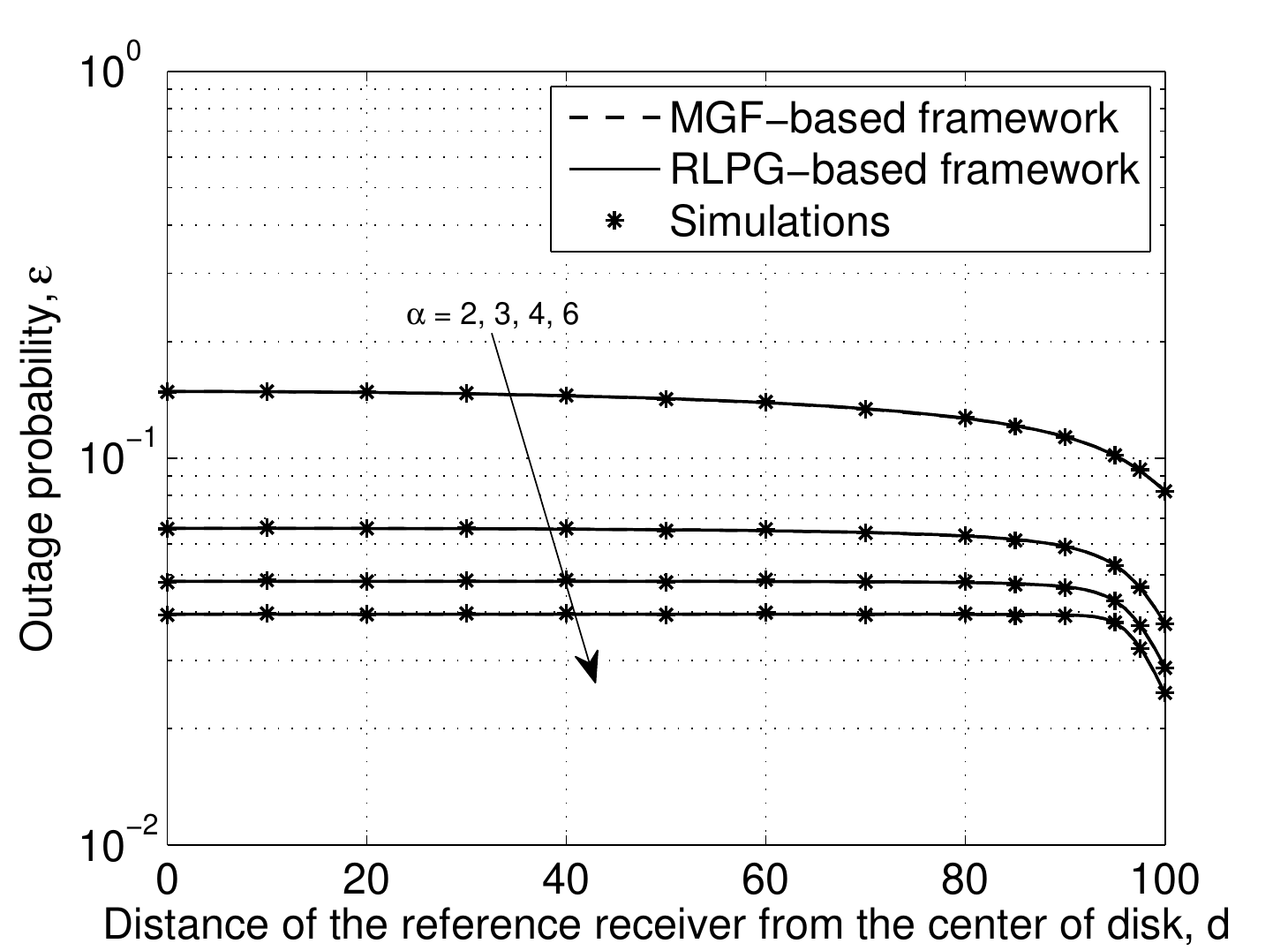}
        \caption{Outage probability, $\epsilon$, versus the distance of the reference receiver from the center of disk, $d$, for $M=10$ interferers i.u.d. in a disk of radius $\radius=100$, with reference link distance $r_0=5$, path-loss exponents $\alpha=2, 3, 4, 6$, i.i.d. Rayleigh fading channels ($m_0=m=1$), SINR threshold $\beta=0$ dB and SNR $\rho_0=20$ dB.}
        \label{disk_fig1}
\end{figure}
\fi

Fig.~\ref{disk_fig1} plots the outage probability, $\epsilon$, versus the distance of the reference receiver from the center of disk, $d$, for path-loss exponents $\alpha=2, 3, 4, 6$ and i.i.d. Rayleigh fading channels. The solid lines are plotted using~\eqref{ana_gamma_outage} and~\eqref{e_disk}, i.e., the RLPG-based framework. The dash lines are plotted using~\eqref{final_gama_outage} and~\eqref{mgf_disk_expectation}, i.e., the MGF-based framework. For the simulation results, we uniformly distribute the users inside a disk region and average the results over $1$ million simulation runs. We can see that the results from both the frameworks are the same and the curves overlap perfectly. In addition, we can see that the simulation results match perfectly with the our analytical results, which is to be expected since we are evaluating the outage probability exactly. These comparisons verify the accuracy of the proposed frameworks.

\subsection{Importance of Having the Two Frameworks}\label{importance}
As stated earlier in Remark~\ref{remark2}, while both $m_0$ and $m$ can take any (integer or non-integer) value in the MGF-based framework, $m_0$ is constrained to take integer values only (while $m$ can take any value) in the RLPG-based framework. It is important to note that for small $m_0$, using the RLPG-based framework and interpolation for non-integer $m_0$ values either does not work or cannot provide accurate approximation results. This is illustrated in Fig.~\ref{disk_fig2} which plots the outage probability, $\epsilon$, versus the SNR, $\rho_0$, for i.i.d. Nakagami-m fading channels ($m_0=m=0.5, 1, 1.5, 2$), path-loss exponent $\alpha=2.5$ and the reference receiver located at the center of the disk. The results for $m_0=m=0.5$ and $1.5$ are plotted using the MGF-based framework (\eqref{final_gama_outage} and \eqref{mgf_disk_expectation}). The results for $m_0=m=1$ and $2$ are  plotted using the  RLPG-based framework (\eqref{ana_disk_center}). For $m_0=m=1.5$ we also plot the arithmetic and the geometric means using the outage probabilities for $m_0 = m = 1$ and $m_0 = m = 2$, respectively. We can see that the arithmetic and geometric means do not match the exact value of the outage probability, which illustrates that the interpolation approach~\cite{abu1991,tellambura1999} does not work here. In addition, the result for $m_0 = m = 1$ does not provide a tight bound on the outage probability when $m_0 = m <1$. These issues highlight the importance of having the two frameworks, which together can handle any value of $m_0$.

Note that the error floor observed in all the curves in Fig.~\ref{disk_fig2} is due to the fact that at high SNR, the interference term dominates the noise term and causes the outage probability to become nearly constant (the $x-$axis in Fig.~\ref{disk_fig2} is the SNR, which is defined below~\eqref{sinr_define} and not the SINR, which is defined in~\eqref{sinr_define}).

%
\ifCLASSOPTIONonecolumn
\begin{figure}[t]
\centering
        \includegraphics[width=0.65  \textwidth]{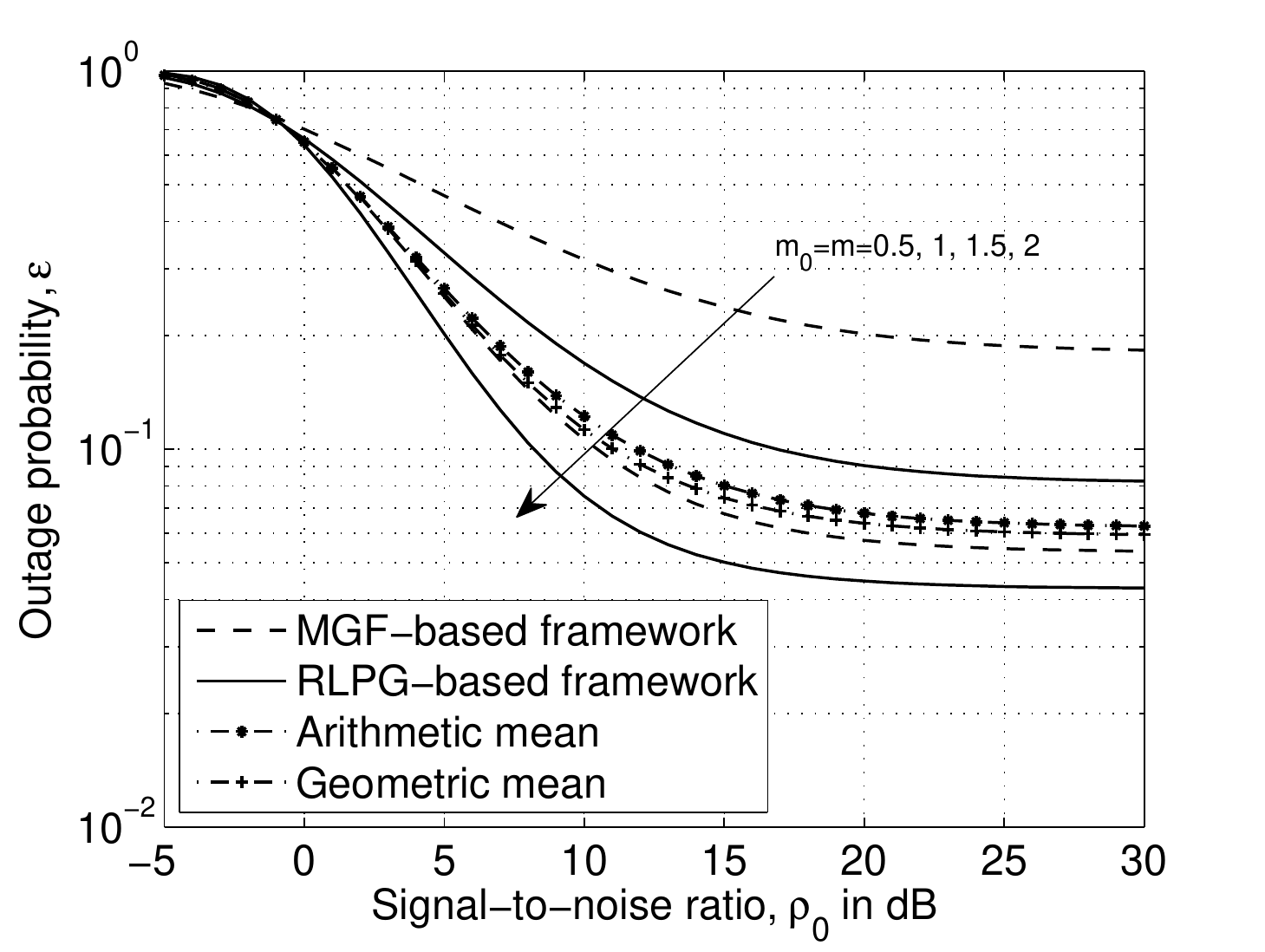}
        \caption{Outage probability, $\epsilon$, versus the signal-to-noise ratio, $\rho_0$, for i.i.d. Nakagami-$m$ fading channels and $m_0=m=0.5, 1, 1.5, 2$, with $M=10$ interferers i.u.d. in a disk of radius $\radius=100$, reference link distance $r_0=5$ and reference receiver located at the center of the network, path-loss exponent $\alpha=2.5$ and SINR threshold $\beta=0$ dB.}
        \label{disk_fig2}
\end{figure}
\else
\begin{figure}[t]
\centering
        \includegraphics[width=0.5 \textwidth]{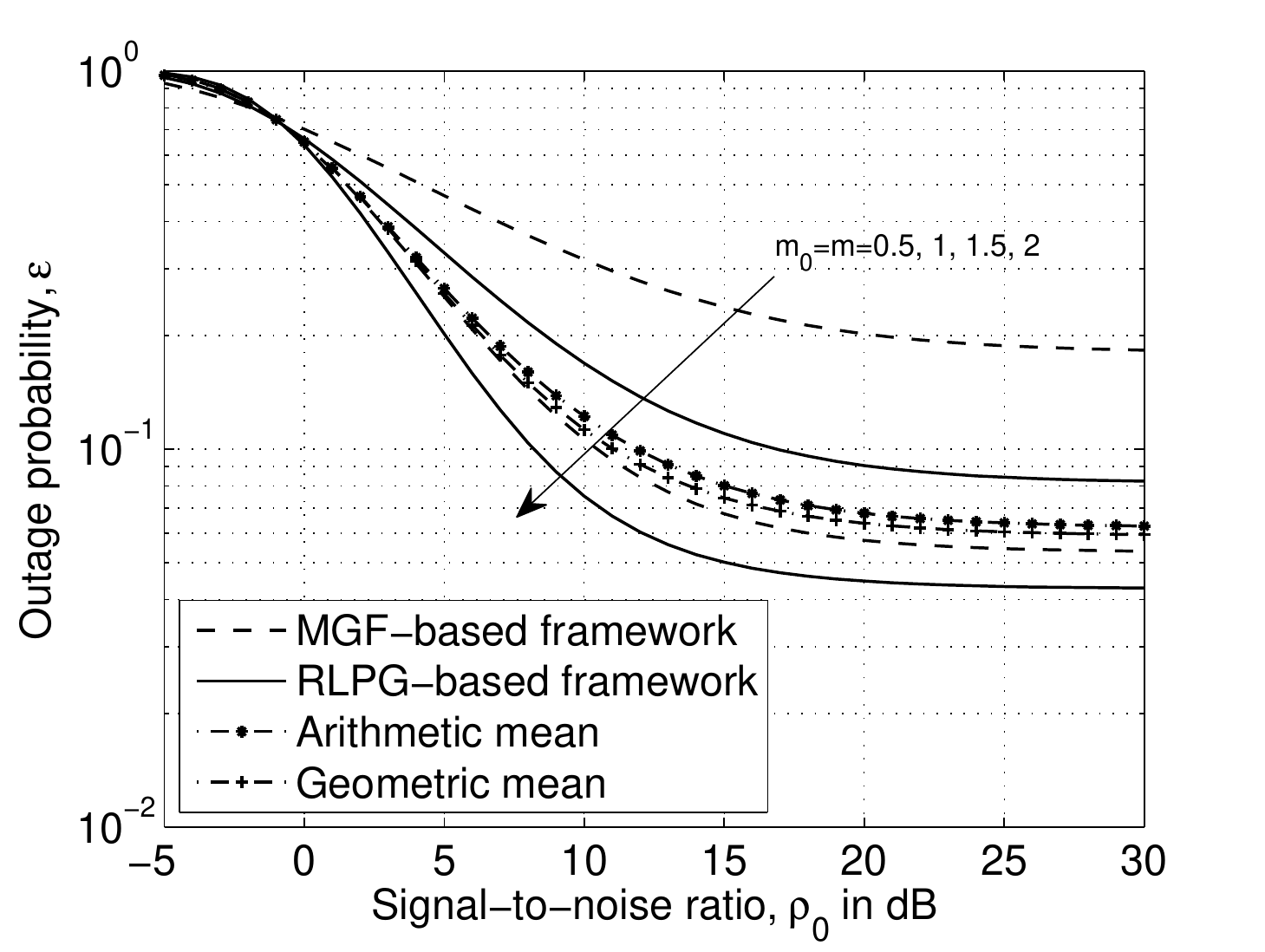}
        \caption{Outage probability, $\epsilon$, versus the signal-to-noise ratio, $\rho_0$, for i.i.d. Nakagami-$m$ fading channels and $m_0=m=0.5, 1, 1.5, 2$, with $M=10$ interferers i.u.d. in a disk of radius $\radius=100$, reference link distance $r_0=5$ and reference receiver located at the center of the network, path-loss exponent $\alpha=2.5$ and SINR threshold $\beta=0$ dB.}
        \label{disk_fig2}
\end{figure}
\fi

\subsection{Boundary Effects in a Disk Region}
Fig.~\ref{disk_fig1} shows that for the disk region the minimum value of the outage probability occurs when the reference receiver is located at the circumference. This is due to the boundary effects. When the nodes are confined within a finite region, the nodes located close to the physical boundaries of the region experience different network characteristics, such as outage probability, compared to the nodes located near the center of the region. Note that the boundary effects are absent in PPP networks which assume an infinite region. In the following, we focus on the low outage probability regime and study the impact of the system parameters on the boundary effects by comparing the performance at the center and the boundary of disk and polygon regions, respectively.
\ifCLASSOPTIONonecolumn
\begin{figure}[t]
\centering
        \includegraphics[width=0.65 \textwidth]{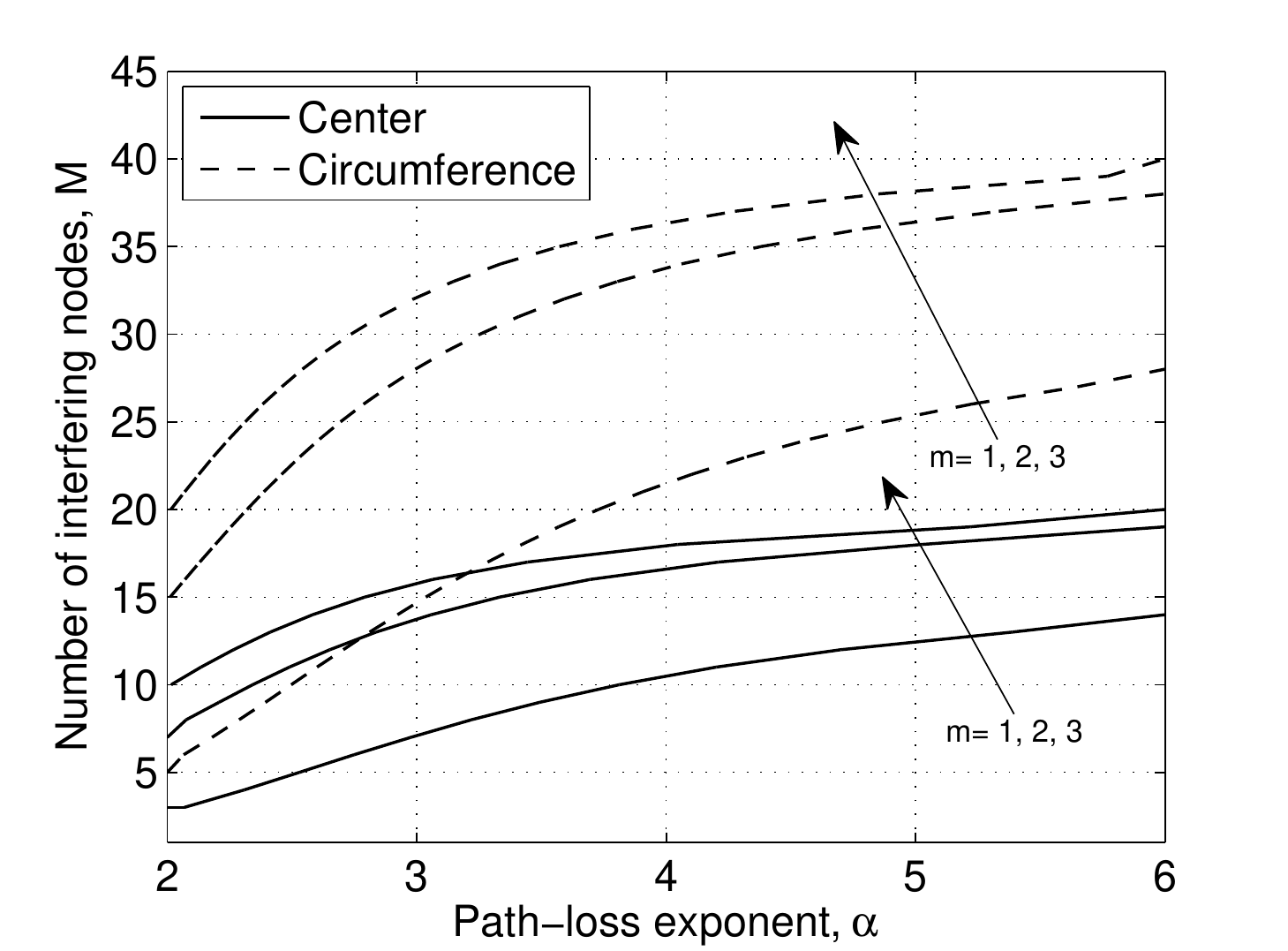}
        \caption{The number of interfering nodes $M$ versus path-loss exponent $\alpha$ in order to meet a fixed outage probability constraint of $\epsilon=0.05$ for the reference node located at the center and the circumference, respectively, of a disk region with radius $\radius=100$, i.i.d. Nakagami-$m$ fading channels ($m =1, 2, 3$), reference link distance $r_0=5$, SINR threshold $\beta =0$ dB and SNR $\rho_0=20$ dB.}
        \label{fig3d_disk}
\end{figure}
\else
\begin{figure}[t]
\centering
        \includegraphics[width=0.5 \textwidth]{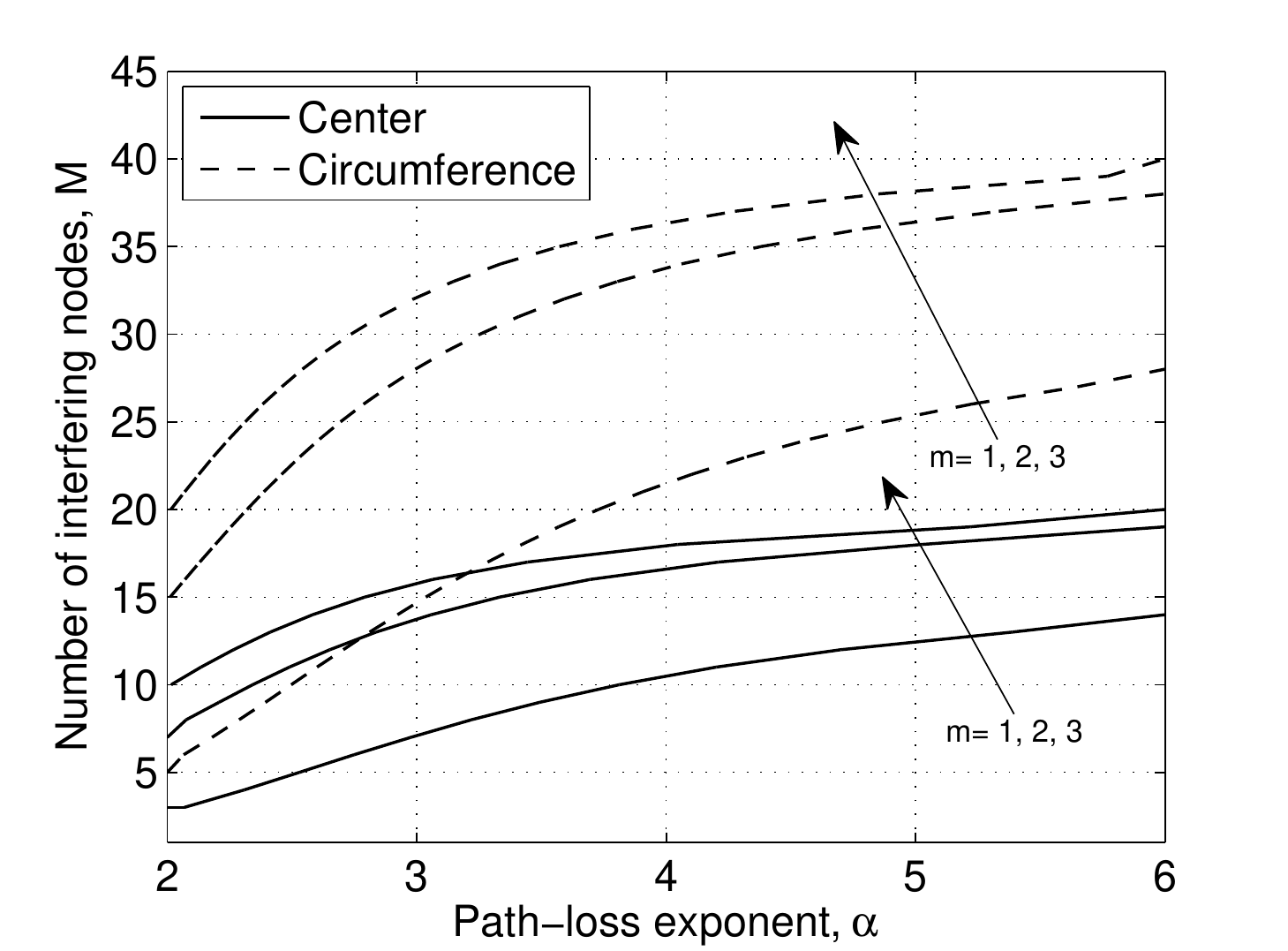}
        \caption{The number of interfering nodes $M$ versus path-loss exponent $\alpha$ in order to meet a fixed low outage probability constraint of $\epsilon=0.05$ for the reference node located at the center and the circumference, respectively, of a disk region with radius $\radius=100$, i.i.d. Nakagami-$m$ fading channels ($m =1, 2, 3$), reference link distance $r_0=5$, SINR threshold $\beta =0$ dB and SNR $\rho_0=20$ dB.}
        \label{fig3d_disk}
\end{figure}
\fi

Fig.~\ref{fig3d_disk} plots the number of interfering nodes, $M$, that the network can accommodate in order to meet a fixed low outage probability constraint of $\epsilon=0.05$ versus the path-loss exponent $\alpha$ for the two cases of the reference receiver located at the center and the circumference of a disk region, respectively, with radius $\radius=100$, i.i.d. Nakagami-$m$ fading channels ($m =1, 2, 3$), reference link distance $r_0=5$, SINR threshold $\beta =0$ dB and SNR $\rho_0=20$ dB. We can see that as the path-loss exponent $\alpha$ increases the number of interfering nodes increases for all the curves. This is because as $\alpha$ increases, the total received power at the reference receiver $Y_0$ from all the interferers decreases more as compared to the received power at $Y_0$ from the desired transmitter $X_0$. In addition, as $m_0=m$ increases, the number of interfering nodes increases for all the curves. This is because as the the fading becomes less severe, the received power at the reference receiver $Y_0$ from the desired transmitter $X_0$ increases more compared to the total received power from all the interferers. Comparing the curves for the center and the circumference, we can see that when $Y_0$ is located at the center of the disk region the network can only accommodate a small number of interferers in order to meet the low outage constraint. However, when $Y_0$ is located at the circumference of the disk region, the network can accommodate a larger number of interferers. This is because the circumference location is most impacted by the boundary effects and the reference receiver located at the circumference can only experience interference from certain surrounding regions inside the disk region. The figure also shows that the two sets of curves for the center and the circumference are not parallel, i.e., the impact of the boundary effects varies with the channel conditions. This is further explored in the next figure.

Fig.~\ref{diff3d_disk} plots the difference in the number of interferers between the reference receiver located at the circumference and the center of a disk region, versus the path-loss exponent $\alpha$ for the scenario considered in Fig.~\ref{fig3d_disk}. The figure shows that for a fixed $m_0=m$, an increase in the path-loss exponent enhances the impact of the boundary effects, e.g., for $m_0= m=1$ the difference is $2$ interferers for $\alpha=2$, which grows to $14$ interferers for $\alpha=6$. In addition, for a fixed $\alpha$, an increase in $m_0=m$ also enhances the impact of the boundary effects, e.g., for $\alpha =4$, the difference is $11$ interferers for $m_0=m=1$, which grows to $18$ interferers for $m_0=m=3$. Note that the staircase nature of the curves in Fig.~\ref{diff3d_disk} is due to the fact that the difference in the number of the interferers can only take integer values.

\ifCLASSOPTIONonecolumn
\begin{figure}[t]
\centering
        \includegraphics[width=0.65 \textwidth]{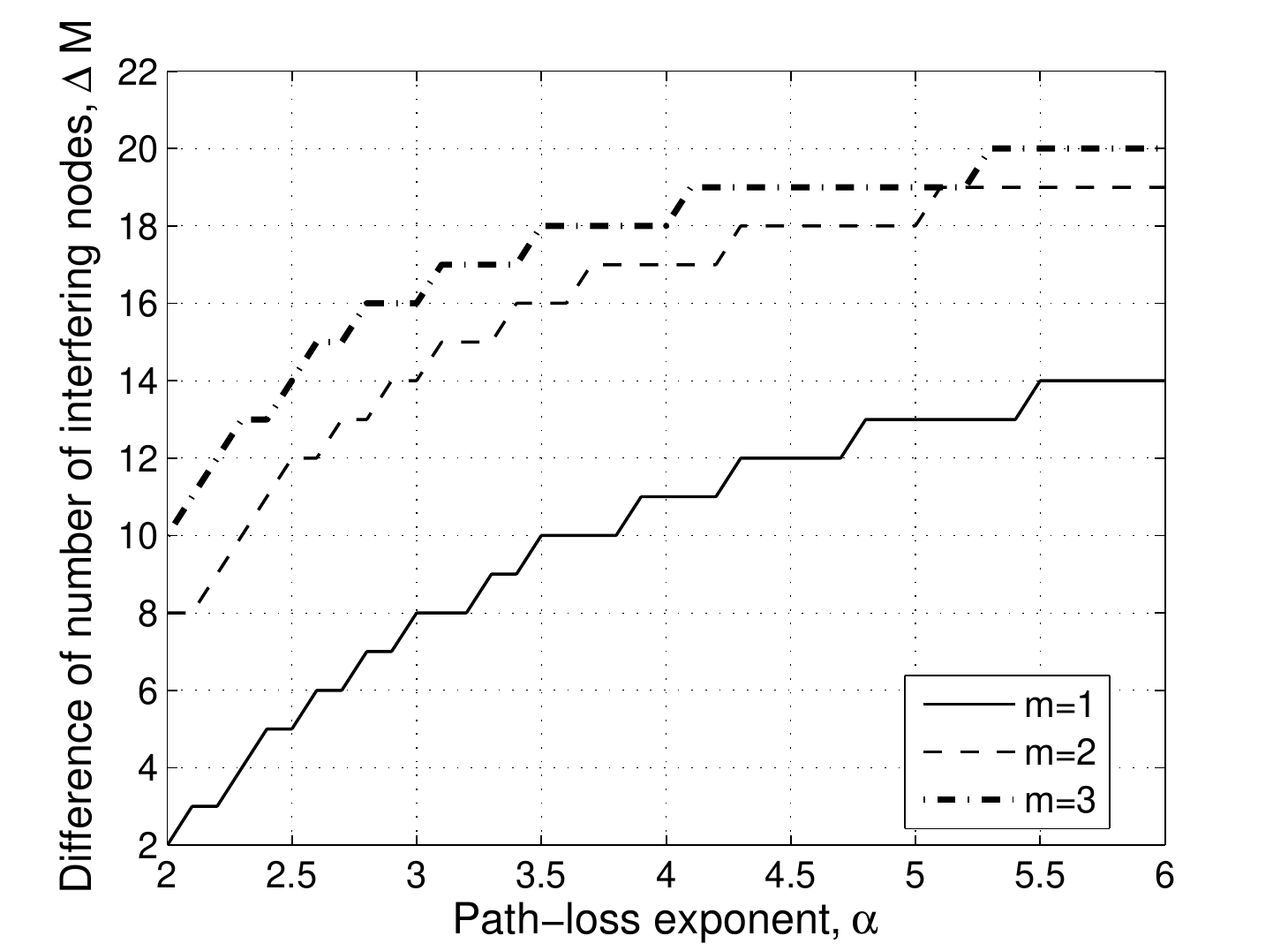}
   \caption{The difference in the number of interferers between the reference receiver located at the circumference and the center of a disk region, versus the path-loss exponent $\alpha$, with the same parameters as in Fig.~\ref{fig3d_disk}.}
        \label{diff3d_disk}
\end{figure}
\else
\begin{figure}[t]
\centering
        \includegraphics[width=0.5 \textwidth]{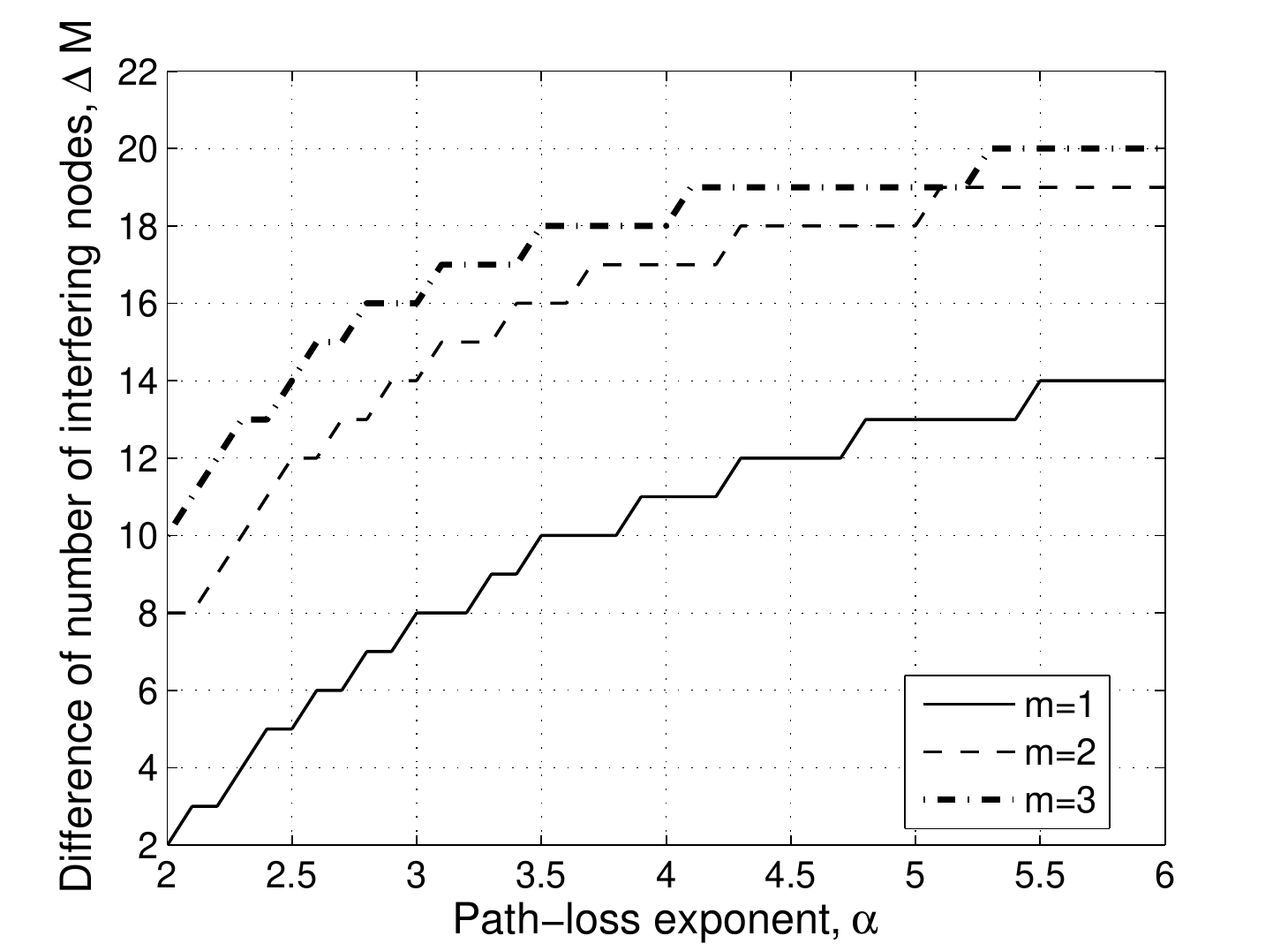}
   \caption{The difference in the number of interferers between the reference receiver located at the circumference and the center of a disk region, versus the path-loss exponent $\alpha$, with the same parameters as in Fig.~\ref{fig3d_disk}.}
        \label{diff3d_disk}
\end{figure}
\fi

\subsection{Boundary Effects in Polygon Regions}
Fig.~\ref{fig3d_polygon} shows the number of interfering nodes $M$ that the network can accommodate in order to meet a fixed low outage probability constraint of $\epsilon=0.05$ versus the number of sides $L$ for the two cases of the reference receiver located at the center and a vertex of a $L=3,4,5,6,7,8,9$-sided convex polygon having a fixed area $|\mathcal{A}|=\pi 100^2$, with i.i.d. Nakagami-$m$ fading channels ($m_0=m =3$), path-loss exponent $\alpha=2.5$, reference link distance $r_0=5$, SINR threshold $\beta =0$ dB and SNR $\rho_0=20$ dB. When the reference receiver is located at the center of the polygon, the impact of the boundary effects is negligible and the number of interferers that the network can accommodate is 14, irrespective of the number of sides{\footnote{For the case of the reference receiver located at the center of a disk region (which can be regarded as a $L=\infty$-sided convex polygon) with the same area, the number of interferers is also 14.}. This is consistent with the fact that we have considered a large-scale finite network and the surrounding environment for the reference receiver located at the center of the $L$-sided convex polygon is quite the same, regardless of the number of sides. When the reference receiver is located at a vertex of the $L$-sided convex polygon, the number of interferers decreases as the number of sides $L$ increases. This shows that the impact of the number of sides on the boundary effects depends on the location of the reference receiver. We can see that for $L=3$ the network can accommodate the highest number of interferers while meeting the low outage probability constraint of $\epsilon=0.05$. This can be intuitively explained as follows. For $L=3$, since the area of the $L$-sided polygon is fixed, $f_R(r)$ has the longest tail. In addition, the interior angle formed at the vertex is the smallest. Consequently, the interfering nodes are more likely to located further away from the reference receiver located at a vertex, which leads to better performance in terms of the number of interferers that the network can accommodate. In addition, Fig.~\ref{fig3d_polygon} shows that as $L$ increases, the difference in the number of interferers between the two cases of the reference receiver located at the vertex and the center decreases. This shows that the relative impact of the boundary effects becomes less significant as the number of sides increases.
%
\ifCLASSOPTIONonecolumn
\begin{figure}[t]
\centering
        \includegraphics[width=0.65 \textwidth]{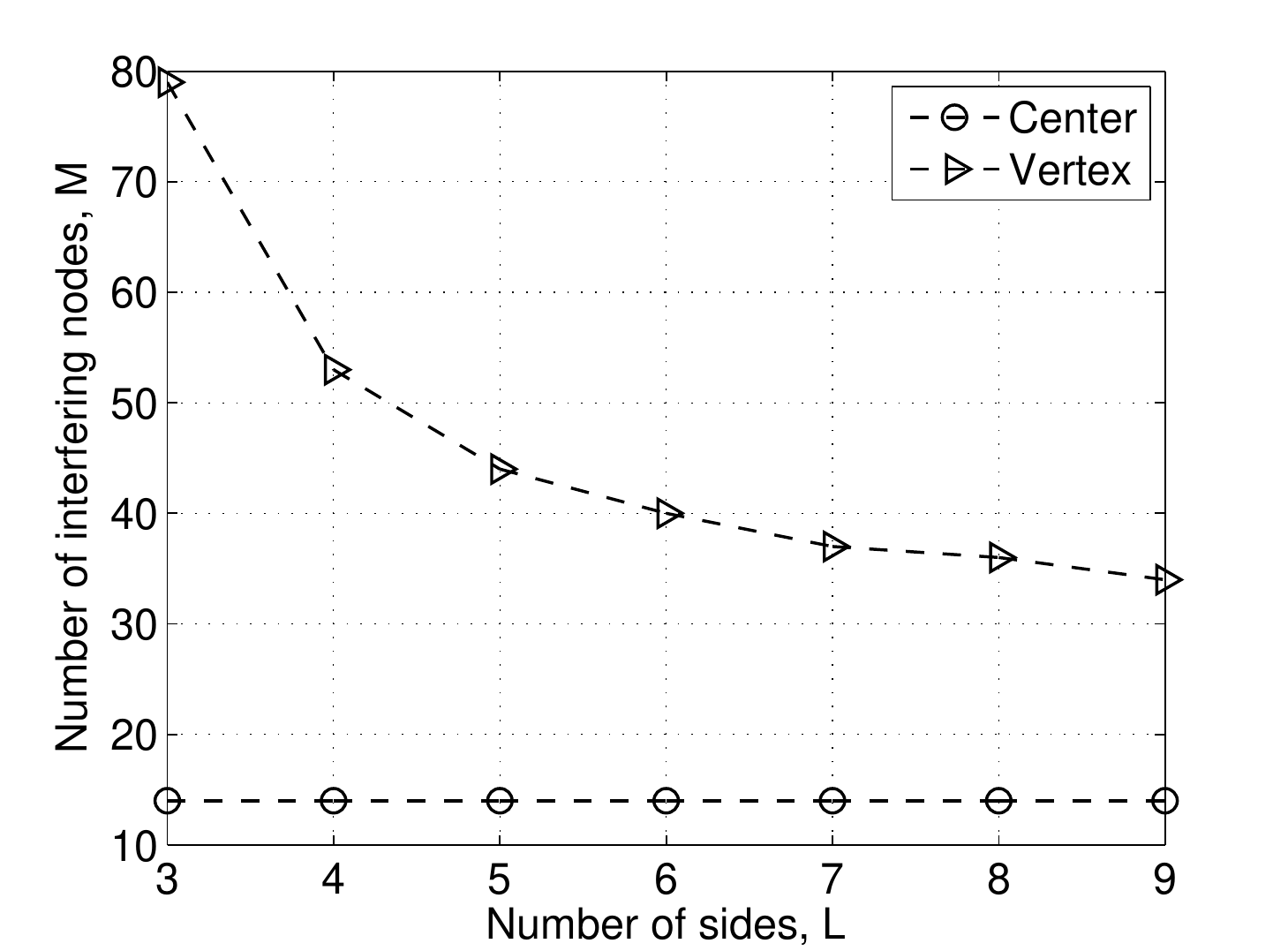}
        \caption{The number of interfering nodes $M$ versus the number of sides $L$ in order to meet a fixed low outage probability constraint of $\epsilon=0.05$ for the reference node located at the center and the corner, respectively, of a $L$-sided polygon ($L=3,4,5,6,7,8,9$) having a fixed area $|\mathcal{A}|=\pi 100^2$, with i.i.d. Nakagami-$m$ fading channels ($m_0= m=3$), path-loss exponent $\alpha=2.5$, reference link distance $r_0=5$, SINR threshold $\beta=0$ dB and SNR $\rho_0=20$ dB.}
        \label{fig3d_polygon}
\end{figure}
\else
\begin{figure}[t]
\centering
        \includegraphics[width=0.5 \textwidth]{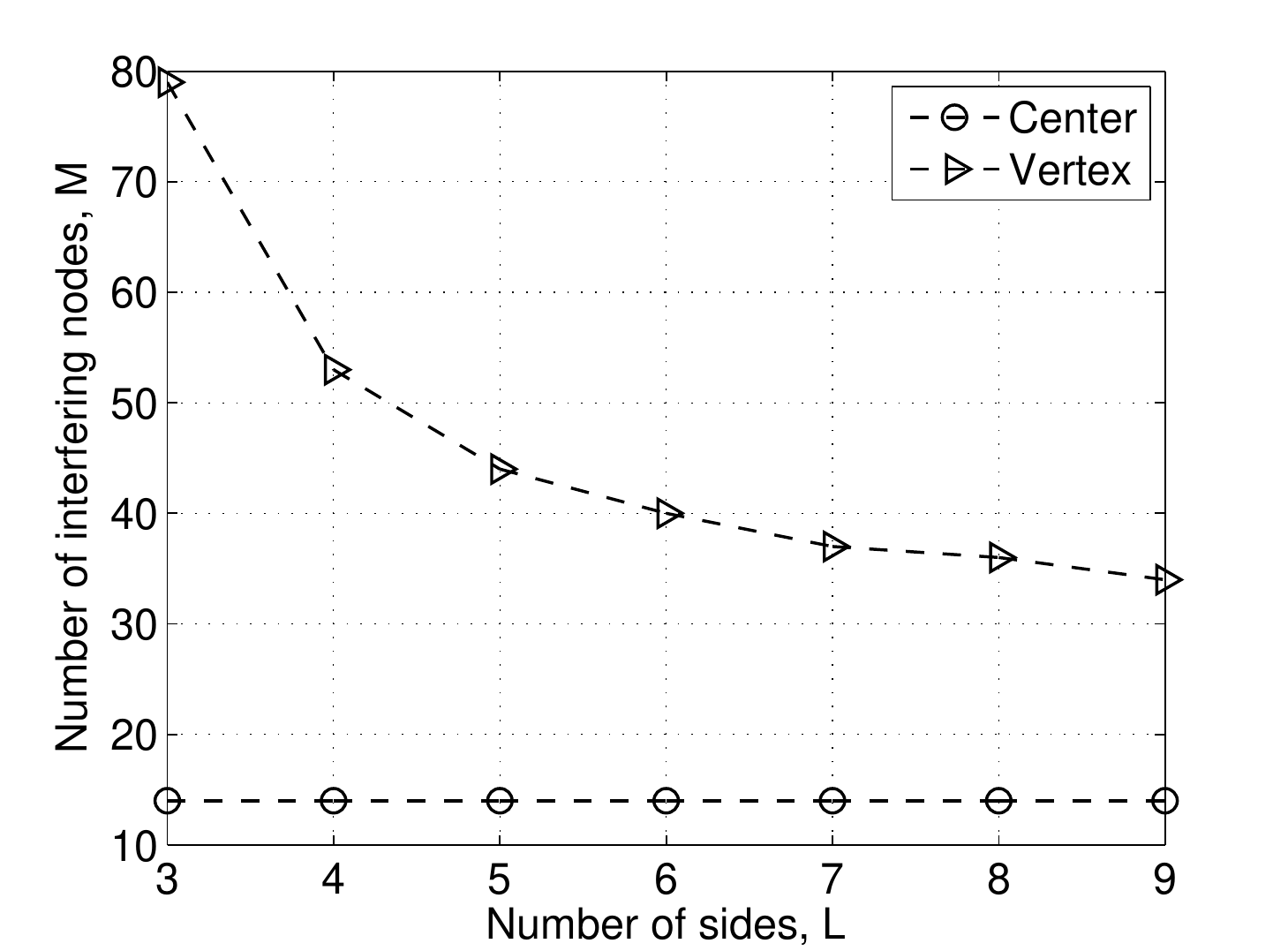}
        \caption{The number of interfering nodes $M$ versus the number of sides $L$ in order to meet a fixed low outage probability constraint of $\epsilon=0.05$ for the reference node located at the center and the corner, respectively, of a $L$-sided polygon ($L=3,4,5,6,7,8,9$) having a fixed area $|\mathcal{A}|=\pi 100^2$, with i.i.d. Nakagami-$m$ fading channels ($m_0= m=3$), path-loss exponent $\alpha=2.5$, reference link distance $r_0=5$, SINR threshold $\beta=0$ dB and SNR $\rho_0=20$ dB.}
        \label{fig3d_polygon}
\end{figure}
\fi


\subsection{Outage Probability in an Arbitrarily-shaped Convex Region}
Fig.~\ref{arb_fig1} plots the outage probability, $\epsilon$, versus the SNR, $\rho_0$, with arbitrary locations of the reference receiver in the arbitrarily-shaped finite region ($|\mathcal{A}|=13143$) defined in Fig.~2, i.i.d. Rayleigh fading channels ($m_0=m=1$) and path-loss exponent $\alpha=2.5$. We consider and compare the following cases for the reference receiver located at: (i) vertex $V_2$ at ($173.2,0$) (ii) vertex $V_3$ at ($50.73,122.474$) (iii) mid point of side $S_2$ at ($111.97,61.24$) and (iv) intersection point of the diagonals at ($33.4,80.7$). For comparison, we also plot the outage probability assuming a PPP node distribution with a node density $\lambda= 10/13143=7.6086\times10^{-4}$ using the result from~\cite{Weber-2010}, which is given below
\begin{align}
\epsilon=1-\exp\left(-\frac{\beta}{\rho_0}\right)\exp\left(-\lambda\pi r_0^2\beta^{\frac{2}{\alpha}}\frac{2\pi}{\alpha}\csc\left(\frac{2\pi}{\alpha}\right)\right).
\end{align}

At high SNR, the error floor observed in all the curves is because of the same reasons as explained before. We can see that the outage probabilities for the four cases are completely different as the location of the reference receiver and consequently the boundary effects are different in each case. The outage probability is the highest for case (iv) as this location is well inside the region and is less impacted by the boundary effects. The outage probability is the lowest for case (i) as the interior angle formed at $V_2$ vertex is the smaller than that at vertex $V_3$. Thus, interferers are more likely to be located further away from the reference receiver located at $V_2$ vertex. We can see that the PPP result, which does not take boundary effects into account, is completely different from the four cases considered and provides an extremely loose upper bound for the outage probability. This re-iterates the importance of the proposed frameworks, which allow the outage probability at any arbitrary location of a finite wireless network with arbitrary shape to be exactly determined.

%
\ifCLASSOPTIONonecolumn
\begin{figure}[t]
\centering
    \includegraphics[width=0.65\textwidth]{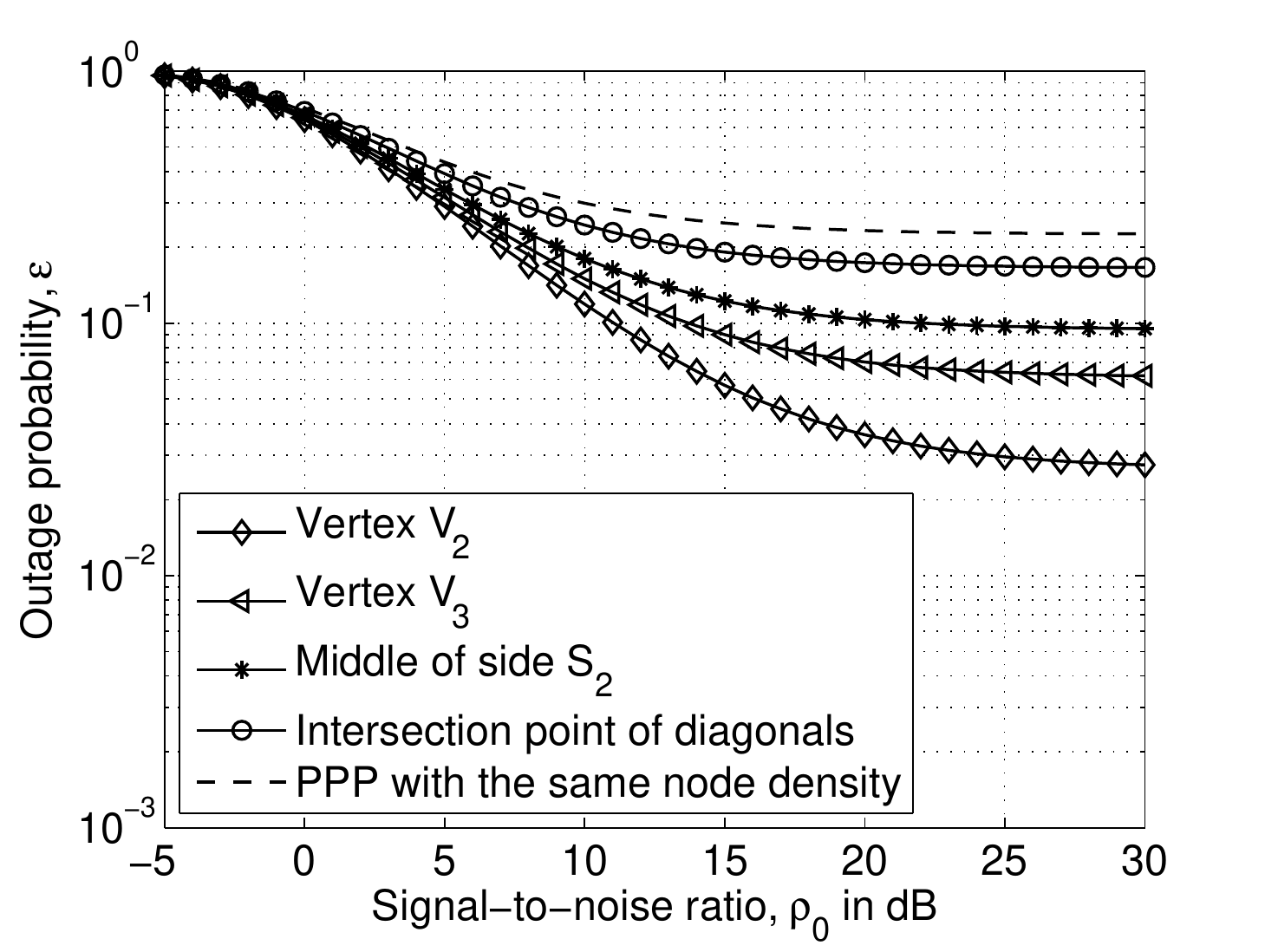}
   \caption{Outage probability, $\epsilon$, versus the signal-to-noise ratio, $\rho_0$, with arbitrary locations of the reference receiver inside the arbitrarily-shaped finite region defined in Fig.~2 having area $|\mathcal{A}|=13143$, $M=10$ interferers, i.i.d. Rayleigh fading channels ($m_0=m=1$), path-loss exponent $\alpha=2.5$ and SINR threshold $\beta=0$ dB. For the PPP node distribution, the node density is $\lambda= 10/13143= 7.6086\times10^{-4}$ (which is the same as for the BPP).}  \label{arb_fig1}
\end{figure}
\else
\begin{figure}[t]
        \centering
    \includegraphics[width=0.5\textwidth]{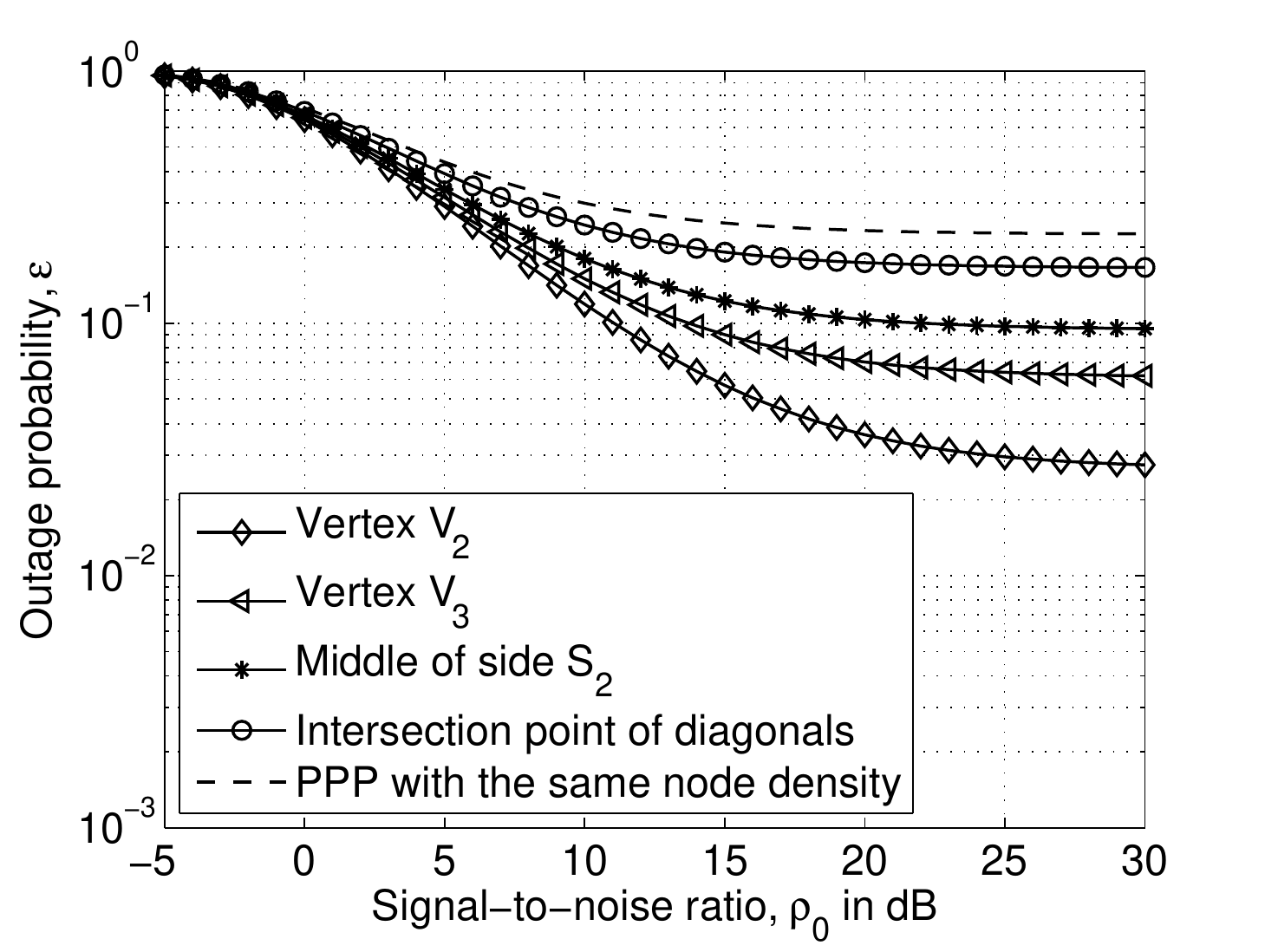}
   \caption{Outage probability, $\epsilon$, versus the signal-to-noise ratio, $\rho_0$, with arbitrary locations of the reference receiver inside the arbitrarily-shaped finite region defined in Fig.~2 having area $|\mathcal{A}|=13143$, $M=10$ interferers, i.i.d. Rayleigh fading channels ($m_0=m=1$), path-loss exponent $\alpha=2.5$ and SINR threshold $\beta=0$ dB. For the PPP node distribution, the node density is $\lambda= 10/13143= 7.6086\times10^{-4}$ (which is the same as for the BPP).} \label{arb_fig1}
\end{figure}
\fi

\section{Conclusion}\label{conclusion}

In this paper, we have proposed two general frameworks for analytically computing the outage probability when a finite number of nodes are distributed at random inside an arbitrarily-shaped finite wireless network. For the case that the random nodes are independently and uniformly distributed and the fading channels are identically, independently distributed following the Nakagami-$m$ distribution, we have demonstrated the use of the frameworks to analytically compute the outage probability at any arbitrary location inside the finite wireless network. The probability distribution function of a random node from the reference receiver plays a key part in our frameworks and enables the accurate modeling of the boundary effects. We have presented an algorithm to accurately compute the outage probability in a finite wireless network with an arbitrary shape. We have also analyzed the impact of the fading channel and the shape of the region on the boundary effects.

\appendices
\section{Derivation of the Distance Distribution $\fr$}\label{appendix:fr}

In this appendix, we summarize the derivation of the distance distribution results using the procedure in~\cite{Zubair-2013}.

For the uniform BPP, the CDF $F_R(r)$ of the distance between a randomly located node and an arbitrary reference point located inside a regular convex polygon was derived in~\cite{Zubair-2013}. Using this result, the PDF of the Euclidean distance between the arbitrary reference point and its $i$th neighbor node is obtained, generalizing the result in~\cite{Srinivasa-2010}. The approach in~\cite{Zubair-2013} is also applicable for arbitrarily-shaped convex polygons. In this paper, we adapt the procedure in~\cite{Zubair-2013} to obtain the PDF $\fr$ for an arbitrary location of the reference receiver inside an arbitrarily-shaped finite wireless network. The main steps of the procedure are summarized below.

\underline{Step 1:} Consider a convex\footnote{All the interior angles are less than $\pi$ radians.} polygon with $L$ sides. Let $S_\ell$ and $V_\ell$ $(\ell=1,2,\hdots,L)$ denote the sides and the vertices, which are numbered in an anti-clockwise direction.

\underline{Step 2:} Using the property of uniform BPP the CDF $F_R(r)$, which is the probability that the distance between an i.u.d. node and the reference receiver is less than or equal to $r$, is given by
\begin{align}\label{appendix_Fr}
F_R(r)=\frac{\left|\mathcal{D}(Y_0;r)\cap \area\right|}{|\area|}=\frac{\mathcal{O}(Y_0;r)}{|\area|},
\end{align}

\noindent where $\mathcal{D}(Y_0;r)$ is a disk region centered at $Y_0$ with radius $r$ and $\mathcal{O}(Y_0;r)$ is the overlap area between $\mathcal{D}(Y_0;r)$ and the network region, respectively.

\underline{Step 3:} In order to find the area of overlap region $\mathcal{O}(Y_0;r)$, the approach in~\cite{Zubair-2013} is to first find the circular segment areas formed outside the sides (denoted as $B_\ell$ for side $\ell$) and the corner overlap areas between two circular segments at a vertex (denoted as $C_\ell$ for vertex $\ell$), and then subtract from the area of the disk. Thus,~\eqref{appendix_Fr} can be expressed as
\begin{align}\label{appendix_Fr_mod}
F_R(r)=\frac{1}{|\area|}\left(\pi r^2 - \sum_\ell B_\ell +\sum_\ell C_\ell \right).
\end{align}

Taking the derivative of~\eqref{appendix_Fr_mod}, we have
\begin{align}\label{appendix_fr}
f_R(r)=\frac{1}{|\area|}\left(2\pi r - \sum_\ell \frac{\partial B_\ell}{\partial r} +\sum_\ell \frac{\partial C_\ell}{\partial r}\right),
\end{align}

\noindent where $\partial/\partial r$ denotes the derivative. Note that $\frac{\partial B_\ell}{\partial r}$ and $\frac{\partial C_\ell}{\partial r}$ depend on the radius $r$ and location of $Y_0$. Following~\cite{Zubair-2013}, they can be expressed as
\begin{align}\label{b_form}
\frac{\partial B_\ell}{\partial r}=2 r \arccos\left(\frac{p_{S_\ell}}{r}\right),
\end{align}
\begin{align}\label{c_form}
\frac{\partial C_\ell}{\partial r}=r \left(-\pi + \delta_\ell + \arccos\left(\frac{p_{S_\ell}}{r}\right)+ \arccos\left(\frac{p_{S_{\ell-1}}}{r}\right) \right),
\end{align}

\noindent where $p_{S_\ell}$ denotes the perpendicular distance from $Y_0$ to the side $S_\ell$, $\delta_\ell$ represents the interior angle at vertex $V_\ell$ and the notation $S_\ell$. Note that for $\ell=1$, $S_{\ell-1}=S_{L}$.

\underline{Step 4:} In order to find the perpendicular distances $p_{S_\ell}$, establish a Cartesian coordinate system. For arbitrarily-shaped finite regions, without loss of generality, the origin can be placed at vertex $V_1$. For regular convex polygons, the origin can be placed at the center of the polygon to exploit the rotational symmetry~\cite{Zubair-2013}.

Following the steps above, and substituting the values, the distance distribution $\fr$ can be derived. This is illustrated in detail in the next subsection for the case of the reference receiver located at vertex $V_2$ in Fig.~\ref{fig:two}. The remaining distance distribution results used in Section~\ref{result} can be similarly derived, but are not included here for brevity.

\subsection{Reference Receiver Located at Vertex $V_2$ in Fig.~\ref{fig:two}}

Consider the arbitrarily-shaped region shown in Fig.~\ref{fig:two}, with reference receiver located at vertex $V_2$.

For $0 \leq r \leq \sqrt{3}\radius$, the disk region $\mathcal{D}(Y_0;r)$ is limited by sides $S_1$ and $S_2$ and vertex $V_2$, i.e., there are two circular segment areas outside $S_1$ and $S_2$, which are denoted by $B_1$ and $B_2$. Also there is a corner overlap area between them, which is denoted by $C_2$. Hence for this range, $\fr=\frac{1}{|\area|}\left(2\pi r -\frac{\partial B_1}{\partial r}-\frac{\partial B_2}{\partial r} + \frac{\partial C_2}{\partial r}\right)$.

For $\sqrt{3}\radius \leq r \leq 2\radius$, the disk region $\mathcal{D}(Y_0;r)$ is limited by all four sides and three vertices $V_1$, $V_2$ and $V_3$, i.e., there are four circular segments formed outside all sides and three corner overlaps between them. Hence for this range, $\fr=\frac{1}{|\area|}\left(2\pi r - \sum\limits_{\ell=1}\limits^{4} \frac{\partial B_\ell}{\partial r} +\sum\limits_{\ell=1}\limits^{3} \frac{\partial C_i}{\partial r}\right)$.

For $r > 2\radius$, the disk region covers the whole region in Fig.~\ref{fig:two} which results in $F_R(r)=1$ and $\fr=0$. Consequently, in this case, $\fr$ is a piece-wise function with two ranges only. Using~\eqref{b_form} and~\eqref{c_form}, and simplifying, we get~\eqref{arb_shape_pdf} which is used in the generation of the results in Fig.~\ref{arb_fig1}.


\end{document}